\documentclass[journal=jacsat,manuscript=article]{achemso}

\usepackage{chemformula} 
\usepackage[T1]{fontenc} 

\SectionNumbersOn
\usepackage{hyperref}
\usepackage{url}
\hypersetup{
    colorlinks=true,
    linkcolor=blue,
    filecolor=gray,      
    urlcolor=blue,
    citecolor=blue,
}
\usepackage{breakurl}
\usepackage{bm}
\usepackage{pythonhighlight}
\usepackage{placeins}
\usepackage{graphicx}
\usepackage{layouts}
\usepackage{subfigure}
\usepackage{caption}

\author{Xiaojuan Hu}
\affiliation{Fritz-Haber-Institut der Max-Planck-Gesellschaft, Faradayweg 4-6, 14195 Berlin, Germany}
\email{xhu@fhi-berlin.mpg.de}
\author{Kazi S. Amin}
\affiliation{Centre for Molecular Simulation and Department of Biological Sciences, University of Calgary, 2500 University Drive NW, Calgary, Alberta T2N 1N4, Canada}
\email{kazi.amin@ucalgary.ca}
\author{Markus Schneider}
\affiliation{Fritz-Haber-Institut der Max-Planck-Gesellschaft, Faradayweg 4-6, 14195 Berlin, Germany}
\author{Carmay Lim}
\affiliation{Institute of Biomedical Sciences, Academia Sinica, Taipei 115, Taiwan}
\alsoaffiliation{Department of Chemistry, National Tsing Hua University, Hsinchu 300, Taiwan}
\author{Dennis Salahub}
\affiliation{Centre for Molecular Simulation and Department of Chemistry, University of Calgary, 2500 University Drive NW, Calgary, Alberta T2N 1N4, Canada}
\email{dsalahub@ucalgary.ca}
\author{Carsten Baldauf}
\affiliation{Fritz-Haber-Institut der Max-Planck-Gesellschaft, Faradayweg 4-6, 14195 Berlin, Germany}
\email{baldauf@fhi-berlin.mpg.de}

\title[FFAFFURR]
  {System-specific parameter optimization for non-polarizable and polarizable force-fields}

\begin{document}
\begin{center}
\noindent
We dedicate this manuscript to Sergei Noskov, who initiated this work\\and whose much too early death shook us all.
\end{center}

\newpage

\begin{abstract}
\noindent
The accuracy of classical force-fields (FFs) has been shown to be limited for the simulation of cation-protein systems despite their importance in understanding the processes of life.
  Improvements can result from optimizing the parameters of classical FFs or by extending the FF formulation by terms describing charge transfer and polarization effects.
  In this work, we introduce our implementation of the CTPOL model in OpenMM, which extends the classical additive FF formula by adding charge transfer (CT) and polarization (POL).
  Furthermore, we present an open-source parameterization tool, called FFAFFURR that enables the (system specific) parameterization of OPLS-AA and CTPOL models.
  The performance of our workflow was evaluated by its ability to reproduce quantum chemistry energies and by molecular dynamics simulations of a Zinc-finger protein.
\end{abstract}

\newpage

\tableofcontents

\newpage
\section{Introduction}

Metal ions are essential in biological systems and are involved in physiological functions ranging from maintaining protein structure and stability to directly participating in catalytic activities. \cite{sarkar1987metal} 
Approximately one-third of all proteins contain metal ions. \cite{Peters2010} 
As an abundant cation in the human body, \cite{christianson1991structural} Zinc ions are known to play an important role in enzyme catalysis or protein folding$/$stability. 
In aqueous solutions, Zn$^{2+}$ normally coordinates with six water molecules in an octahedral coordination geometry.
However, in a protein environment, Zn$^{2+}$ is often observed to form a tetrahedral coordination structure with four ligating amino acid residues, \cite{patel2007analysis} commonly His and Cys.
Due to the nature of electrostatic interactions, Zn$^{2+}$ also tends to be close to negatively charged residues such as Asp or Glu.
Zn$^{2+}$ is involved in various biological functions by interacting with these residues.
For example, metallothioneins (MTs) \cite{Babu2014, https://doi.org/10.1002/cbic.200800511} are present in all living organisms and are involved in various diseases. \cite{Capdevila2012,Cherian2004Metallo, Durand2010} 
Under physiological conditions, the four mammalian MT isoforms have Zn$_3$Cys$_9$ clusters and Zn$_4$Cys$_{11}$ clusters in their centers as functional groups. 
Zinc-finger proteins are another well-studied class of Zinc-containing proteins. 
Multiple fingers can combine together to carry out many complex functions, 
such as regulating DNA/RNA transcription,\cite{Miller1985Repetitive,Wolfe2000DNA} 
protein folding and assembly, lipid binding, Zinc sensing,\cite{laity2001zinc} and even protein recognition.\cite{Gamsjaeger2007Sticky}
The most well characterized Zinc-finger proteins feature a binding domain with two Cys and two His residues.
The study of the classical Cys$_2$His$_2$ Zinc-finger structures is crucial for a better understanding of their broader functions.

Molecular dynamics (MD) simulations employing molecular mechanics (MM) are widely used in the study of complex biological processes, such as protein folding, protein dynamics, and enzyme catalysis because of their ability to model systems at atomic scales ranging in sizes from thousands to millions of atoms and time scales of milli-seconds. \cite{Mobley2012, Lemkul2016, Jorgensen2004} 
The majority of current MD studies employ classical force-fields (FFs) such as OPLS-AA, \cite{A.Kaminski2001} AMBER, \cite{doi:10.1002/wcms.1121} CHARMM\cite{Huang2017} and GROMOS. \cite{Reif2013} 
It is a challenge for classical force-field models to describe metal–protein interactions due to the strong local electrostatic field and induction effect, \cite{Li2015,Li2017,Amin2020,doi:10.1002/qua.26369,schneider2018relative, wu2010polarizable} for example, computer simulation of Zinc-containing proteins has been a long-standing challenge that appears hard to tackle without explicit treatment of charge-transfer or polarization.

One approach to improve the accuracy of force-fields is to refine the parameters by fitting the model to more and more accurate experimental data or quantum mechanical (QM) calculations. 
For example, force-matching algorithms \cite{Akin-Ojo2008} were used to fit parameters to reproduce \textit{ab initio} forces. 
Empirical Continuum Correction (ECC) \cite{Duboue-Dijon2020,Martinek2018,LeBreton2020} force-fields scale the charges to implicitly take electronic polarization into account.
Several works \cite{li2015systematic, li2015parameterization} tune the Lennard-Jones (LJ) parameters or use a 12-6-4 LJ-type model to simulate charge-induced dipole interactions. 
These efforts have been successful to some extent, however, reparameterization is often time-consuming and labor-intensive. 
There are a few automatic parameterization tools, for example, CHARMM General force-field (CGenFF), \cite{Vanommeslaeghe2010} LigParGen, \cite{dodda2017ligpargen} and Antechamber. \cite{sousa2012acpype, Wang2004} 
These programs typically generate missing parameters for a given system based on analogies with atom types and the relevant parameters available in the corresponding FF or through parameter estimation algorithms. \cite{vassetti2019assessment} 
However, the accuracy of assigning approximate parameters to a specific system is limited, and parameters already present in a given FF may also have to be optimized.
FFparam \cite{kumar2020ffparam} and ForceBalance \cite{wang2014building} enable the tuning of existing FF parameters. 
All these parameterization tools share a common assumption of transferability, which assumes a set of parameters optimal for small organic molecules for a given atom type can be applied in a wide range of chemical and spatial contexts.
It is well known that the presence of electron donors and acceptors can significantly affect molecular properties by polarization effects. \cite{Jorgensen2007} 
LJ parameters are also sensitive to the local environment \cite{Tkatchenko2012, Tkatchenko2009} and long-range electrodynamic screening. \cite{Gobre2013} 
In this regard, a fundamentally different approach to derive environment-specific or molecule-specific parameters is proposed in references \cite{Horton2019, Cole2016, Grimme2014}. 
However, parameters still remain fixed despite structures and environments changing over the course of, e.g., MD simulations.

Another approach to improve FF accuracy in metalloprotein simulations is to introduce more physics in to the model. 
Including polarization effects is a significant step to improve force-fields. \cite{Borodin2009, Cieplak2009} 
There is growing evidence that polarizable force-fields describe ionic systems more accurately than classical force-fields. 
It has been found that the inclusion of polarization plays an important role in the simulation of ion channels, \cite{Allen2004} enzymatic catalysis, \cite{Boulanger2014} protein-ligand binding affinity \cite{Panel2018} and dynamic properties of proteins. \cite{Li2011}

At present, there are three main groups of polarizable force-fields, fluctuating charge, induced point-dipoles, and Drude oscillator models. \cite{Bedrov2019} 
The fluctuating charge models simulate polarization effects by allowing charge to flow through the molecule until the electronegativities of atoms become equalized, while keeping the total charge unchanged. \cite{Olano2005} 
One drawback of the fluctuating charge model is that it fails to capture out-of-plane polarization of planar or linear chemical groups. 
The fluctuating charge formula can also be used in conjunction with induced point-dipoles as a complementary approach to account for charge transfer (CT). \cite{soniat2012effects}
A notable model is SIBFA (Sum of Interactions Between Fragments \textit{Ab Initio} Computed). \cite{piquemal2007toward} 

The induced point-dipole models describe polarization energy as the interaction between static point charges and induced dipole moments.
Notable induced point-dipole models include OPLS/PFF, \cite{friesner2005modeling} AMBER ff02, \cite{cieplak2001molecular} and AMOEBA. \cite{ponder2003force, ren2003polarizable} 
The performance of the induced point-dipole models strongly depends on the accuracy of polarizability parameters. 

The Drude oscillator model simulates the distortion of the electron density by attaching additional charged particles (the oscillators) to each polarizable atom. 
Despite many successes of the Drude oscillator model, \cite{Li2015, ngo2015quantum, villa2018classical} it may be limited when charge transfer between cation and coordinating ligand atoms is significant, for example, Cys$^-$ coordinated to metal ions. \cite{Dudev2014} 
Ngo \textit{et al.} \cite{Ngo2015} and Dudev \textit{et al.} \cite{Dudev2003} showed that the charge located on the coordinating ligand is significantly perturbed due to the presence of Ca$^{2+}$. 
The effect exists not only in the first coordination shell, but also in the second shell. 
Thus, including the description of charge transfer is critical for the development of next-generation polarizable FFs. 

The CTPOL \cite{Sakharov2005,sakharov2009force} model incorporates charge transfer (CT) and polarization effects (POL) into classical force-fields. 
The inclusion of charge transfer reduces the amount of partial charge on cation and cation coordinating atoms. 
Thus, their charge/dipole–charge interactions are weakened. 
Local polarization energy between cation and coordinating ligands, which also depends on the partial charge, is introduced for compensation. 

Although numerous studies have shown that polarizable models perform better than classical force-fields in the simulation of metalloproteins, they have received only limited validation. 
Therefore, reparameterization may be necessary when applied to different systems. 
Our previous study \cite{Amin2020} has shown how QM data \cite{ropo2016first, hu2022better} drive the parameter development of the Drude and CTPOL models. 
However, most parameterization tools focus on classical force-field models.
FFparam \cite{kumar2020ffparam} provides parameterization of Drude model; a CTPOL parameterization tool is not yet available.


In this work, we fill this gap by (i) implementing the CTPOL model in OpenMM\cite{Eastman2017} and sharing this code \cite{Hu2023CTPOLMD} and (ii) publishing the Framework For Adjusting force-fields Using Regularized Regression (in short FFAFFURR ) an open-source tool, which facilitates the parameterization of OPLS-AA and CTPOL models for a specific system, e.g. a peptide system or a peptide-cation system.
A major advantage of FFAFFURR is the rapid construction of FFs for troublesome metal centers in metalloproteins.
In this work, the new parameters obtained from FFAFFURR were validated by the comparison of FF energies and QM energies in isolation and by assessing the stability of condensed phase MD simulations using a Zinc-finger protein as an example. 

\section{Methods}

\subsection{OPLS-AA functional form}

OPLS-AA is one of the major families of classical force-fields. 
It is used as the starting point for parameterization in this work. 
OPLS-AA uses the harmonic functional form to represent the potential energy shown in eq. \ref{additive_FF}.
\begin{equation}
E^{\mathrm{FF}} =  E_{\mathrm{bonds}} + E_{\mathrm{angles}} + E_{\mathrm{torsions}} + E_{\mathrm{improper}} + E_{\mathrm{vdW}} + E_{\mathrm{ele}} \label{additive_FF} 
\end{equation}
where $E^{\mathrm{FF}}$ is the potential energy of the system. 
$E_{\mathrm{bonds}}$, $E_{\mathrm{angles}}$, $E_{\mathrm{torsions}}$ and $E_{\mathrm{improper}}$ correspond to bonded or so-called covalent terms of bond stretching, bond-angle bending, dihedral-angle torsion, and improper dihedral-angle bending (or out-of-plane distortions) in the molecules.
$E_{\mathrm{vdW}}$ and $E_{\mathrm{ele}}$ are nonbonded terms. 
They describe van der Waals (vdW) and Coulomb (electrostatic) interactions, respectively.

The energy terms in eq. \ref{additive_FF} are depicted in detail in eq. \ref{additive_FF_detail}.

\begin{equation}
\begin{split}
E^{\mathrm{FF}} =  & \sum_{\mathrm{bonds}}^{\mathrm{1-2 atoms}} \frac{1}{2} K_{ij}^{r} \left (r_{ij}-r_{ij}^0 \right )^{2}+\sum_{\mathrm{angles}}^{\mathrm{1-3 atoms}} \frac{1}{2} K_{ij}^{\theta} \left (\theta_{ij}-\theta_{ij}^0 \right )^{2}+\sum_{\mathrm{dihedrals,n}}^{\mathrm{1-4 atoms}} V_n^{ij} \left ( 1 + cos \left( n\phi _{ij}-\phi_{ij}^0\right)\right )\\
& + \sum_{\mathrm{improper}}^{\mathrm{1-4 atoms}} V_{2imp}^{ij} \left ( 1 + cos \left ( 2\phi _{ij}-\phi_{ij}^0 \right) \right )+\sum_{i<j} 4 \varepsilon_{ij} \left [ \left ( \frac{\sigma_{ij}}{r_{ij}} \right)^{12} - \left (\frac{\sigma_{ij}}{r_{ij}} \right)^{6} \right ] f_{ij}+\sum_{i<j} \frac{q_iq_j}{r_{ij}}f_{ij} \label{additive_FF_detail}
\end{split}
\end{equation}


where $K_{ij}^{r}$, $K_{ij}^{\theta}$, $V_n^{ij}$, and $V_{2imp}^{ij}$ are force constants, $r_{ij}^0$ and $\theta_{ij}^0$ are the reference bond length and bond angle, $r_{ij}$, $\theta_{ij}$ and $\phi_{ij}$ are current bond length, bond angle and dihedral angle, respectively, $n$ is the periodicity, $\phi_{ij}^0$ is the phase offset, $\sigma_{ij}$ is the distance at zero energy, $\varepsilon_{ij}$ sets the strength of the interaction, $q_i$ and $q_j$ are the charges of the two particles, and $f_{ij}$ is the scaling factor for short distances (i.e. ``1-4 pairs'') of nonbonded interaction. 
In OPLS-AA, the pairwise LJ parameters $\sigma_{ij}$ and $\varepsilon_{ij}$ are calculated as the geometric mean of those of individual atom types ($\sigma_{i}$ and $\varepsilon_{i}$). 

Classical force-field simulations were performed using OpenMM7, a high performance toolkit for molecular simulations. \cite{Eastman2017} 

\subsection{CTPOL model}

The CTPOL \cite{Sakharov2005, sakharov2009force} model introduces charge transfer and polarization effects into classical force fields. 
Instead of a fixed-charge model, CTPOL takes the charge transfer from a ligand atom $L$ (O, S, N) to a metal cation into account. 
The amount of transferred charge, $\Delta q_{\mathrm{L-Me}}$, is assumed to depend linearly on the inter-atomic distance, $r_{\mathrm{L-Me}}$
\begin{equation}\label{charge_transfer}
\Delta q_{\mathrm{L-Me}}  =  a_L r_{\mathrm{L-Me}} + b_L. 
\end{equation}

where $a_L$ and $b_L$ are parameters to be determined that are specific for pairs of ligand $L$ (O, S, N) and a metal cation.
The parameters $a_L$ and $b_L$ are of opposite sign, so that the magnitude of charge transfer decreases with distance.
The distance at which $\Delta q_{\mathrm{L-Me}}$ becomes $0$ is

\begin{equation}
    r_{\mathrm{L-Me}}^0 = -\frac{b_L}{a_L}
\end{equation}

Beyond this distance, we assume charge transfer to be $0$. 
This approximates real-life charge transfer, which is generally negligible at distances greater than the sum of the vdW radii of atoms $i$ and $j$, $r^{\mathrm{vdW}}_{ij}$.

Thus, charge on ligand atom $L$, $q_L$, can be calculated as
\begin{equation}
q_L  =  q_L^0 + \Delta q_{\mathrm{L-Me}},
\end{equation}
where $q_L^0$ refers to the charge on atom $L$ in a fixed-charge model.

Polarization energy, $E^{\mathrm{pol}}_r$, can be computed as
\begin{equation}
E^{\mathrm{pol}}_r = -\frac{1}{2}\sum_{i} \bm{\mu_i \cdot {E}^0_i},
\end{equation}
where \bm{$\mu_i$} is the induced dipole on atom $i$ and \bm{$E^0_i$} is the electrostatic field produced by the current charge distribution in the system at the polarizable site \textit{i}. 
The summation is over the metal and the metal-bonded residues. 
A cutoff distance $r^{\mathrm{cutoff}}$, which is equal to the sum of the vdW radii of atoms $i$ and $j$ scaled by a parameter $\gamma$ = 0.92, is introduced to avoid unphysically high induced dipoles at close distance. 
If the distance between atom $i$ and $j$, $r^{ij}$, is smaller than $r^{\mathrm{cutoff}}$, we set $r^{ij}$ equal to $r^{\mathrm{cutoff}}$.
The only parameter here is the atomic polarizability:
\begin{equation}\label{pol}
\bm{\mu_i} = \alpha_i \bm{E_i},  
\end{equation}
where $E_i$ is the total electrostatic field on atom $i$ due to the charges and induced dipoles in the system.

In this work, we have implemented the CTPOL model in OpenMM via a python script, which can be found at \url{https://github.com/XiaojuanHu/CTPOL_MD}. \cite{Hu2023CTPOLMD}
This represents a proof-of-concept implementation, which runs on CPUs. Further code optimization and a transfer to GPUs will likely speed up simulations substantially.

\subsection{Reference data set} \label{ref_stuctures}

\begin{figure}[htbp]
\vspace{0.0in}
\begin{center}
\includegraphics[width=0.75\textwidth]{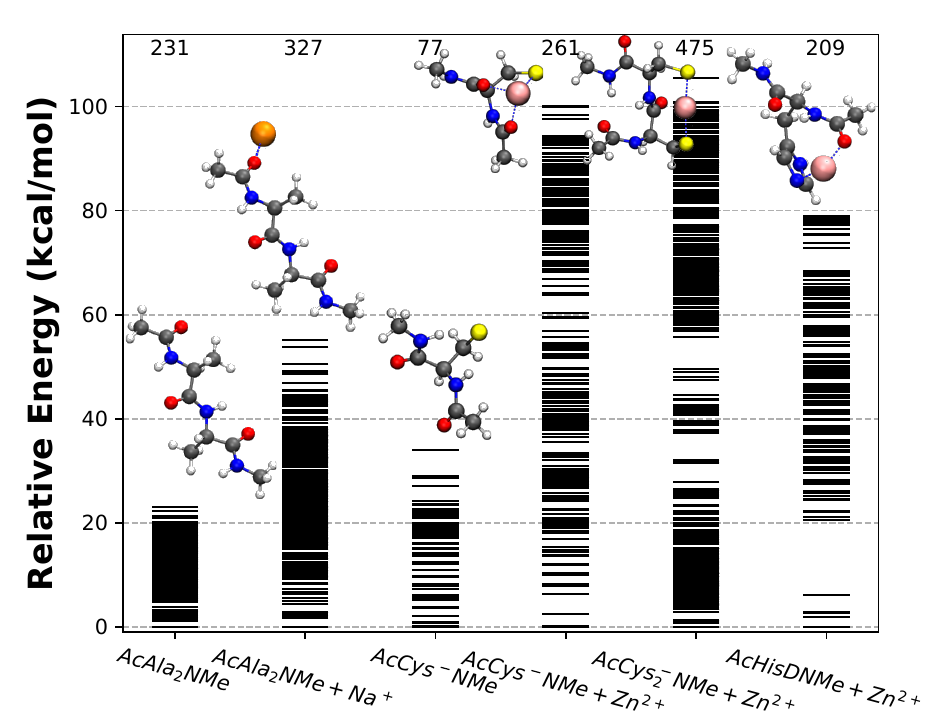}
\end{center}
\vspace{-0.2in}
\setlength{\belowcaptionskip}{0.2in}
\caption{Structures and energy hierarchies of reference data in this study. The numbers of conformers are stated at the top of the individual columns.}
\label{Reference_data}
\vspace{-0.2in}
\end{figure}

To evaluate the performance of the parameterization protocol on dipeptide and dipeptide-cation systems, we created a quantum chemistry data set. 
The data set consists of six models: (1) AcAla$_{2}$NMe (231 conformers); (2) AcAla$_{2}$NMe+Na$^+$ (327 conformers); (3) deprotonated cysteine: AcCys$^-$NMe (77 conformers), which often acts as  interaction center in metalloproteins; (4) AcCys$^-$NMe+Zn$^{2+}$ (261 conformers); (5) AcCys$^{-}_{2}$NMe+Zn$^{2+}$ (475 conformers), and (6) AcHisDNMe+Zn$^{2+}$ (209 conformers). 
The structures and energy hierarchies are shown in Figure \ref{Reference_data}. 
The data set can be found on the NOMAD repository via the DOI: \url{10.17172/NOMAD/2023.02.03-1}. \cite{Hu2023NOMAD}


All DFT calculations in this work were performed with the numeric atom-centered basis set all-electron code FHI-aims. \cite{BLUM20092175, HAVU20098367, Ren_2012}
The PBE \cite{PhysRevLett.77.3865} generalized-gradient exchange-correlation functional augmented by the correction of van der Waals interactions using the Tkatchenko-Scheffler formalism \cite{PhysRevLett.102.073005} (PBE+vdW$^{TS}$) was employed. 
The choice of functional has been validated in previous articles. \cite{ropo2016first, ropo2016trends} 
For each conformation, several types of partial charges were provided. 
Hirshfeld charges \cite{Bultinck2007} are derived based on the Hirshfeld partitioning scheme. \cite{Hirshfeld1977, Bultinck2007} 
ESP charges \cite{Bultinck2007, Singh1984} are derived by fitting partial charges to reproduce the electrostatic potential. 
RESP charges \cite{Bayly1993} are extracted by a two-stage restrained electrostatic potential (RESP) fitting procedure \cite{Bayly1993} within the Antechamber suite of the AmberTools package. \cite{doi:10.1002/wcms.1121} 
The electrostatic potential was evaluated on a set of grids in a fixed spatial region located in a cubic space around the molecule. 
The 5 radial-shells were generated in a radial region between 1.4 and 2.0 multiples of the atomic vdW radius.
The cubic space contains 35 points along $x$, $y$, and $z$ directions, respectively.

The conformers of AcAla$_{2}$NMe, AcAla$_{2}$NMe+Na$^+$, and AcHisDNMe+Zn$^{2+}$ were obtained by a conformational search algorithm as shown in the studies of Rossi \textit{et al.} \cite{Rossi2014} and Schneider \textit{et al}. \cite{schneider2018relative} 
First, a global conformational search was performed with the basin-hopping approach \cite{J.Wales1997, Wales1368} at the force-field level (OPLS-AA). \cite{Jorgensen1996} 
The scan program of the TINKER molecular modeling package \cite{doi:10.1002/jcc.540080710, Ren2003} was employed to perform the basin-hopping search strategy. 
An energy threshold of 100 kcal/mol for local minima and a convergence criterion for local geometry optimizations of 0.0001 kcal/mol were used. 
All obtained conformers were relaxed at PBE+vdW$^{TS}$ level with \textit{tier 1} basis set and \textit{light} setting employed. 
A clustering scheme was then applied to exclude duplicates using the root-mean-square deviations (RMSD) of atomic positions. 
Finally, further relaxation was accomplished at the PBE+vdW$^{TS}$ level using \textit{tier 2} basis set and \textit{tight} setting. 


The conformers of AcCys$^-$NMe, AcCys$^-$NMe+Zn$^{2+}$, and AcCys$^{-}_{2}$NMe+Zn$^{2+}$ were obtained with the genetic algorithm (GA) package Fafoom.\cite{Supady2015} 
First, a GA search at the PBE+vdW$^{TS}$ level with \textit{light} basis set was employed for structure sampling. 
Then a clustering scheme with a clustering criterion of RMSD of 0.02 {\AA} for atomic positions and a relative energy of 0.02 kcal/mol was applied to remove duplicates. 
The obtained conformers were further relaxed with FHI-aims \cite{BLUM20092175, HAVU20098367, Ren_2012} at the PBE+vdW$^{TS}$ level with \textit{tight} basis set. 
Final conformers were obtained after clustering. 
Both conformational search protocols have been well validated. \cite{Rossi2014, Supady2015}




\subsection{Parameter optimization} \label{Method-optimization}

Optimization methods used in this work include LASSO (least absolute shrinkage and selection operator) \cite{Tibshirani1996} regression, Ridge regression \cite{doi:10.1080/00401706.1970.10488634} and particle swarm optimization (PSO). \cite{Shi1945, Koh2007} 
If the parameters enter the force-field function in a quadratic way, e.g. $V_n^{ij}$, the optimization can be performed by solving a set of linear equations.
In this case, LASSO and Ridge regression were employed to treat the potential overfitting. 
The regularization parameter $\lambda$ in LASSO and Ridge regression was selected by 10-fold cross-validation. 
LASSO and Ridge regression were performed with Python’s scikit-learn \cite{pedregosa2011scikit} library. 
If the parameters can not be obtained by solving a set of linear equations, e.g. the charge transfer parameters $a_L$, PSO was employed.
PSO is a powerful population-based global optimization algorithm. 
It relies on a population of candidate solutions, called particles, and finds the optimal solution by moving these particles through a high-dimensional parameter space based on their position and velocity. 
PSO was performed with the python package pyswarm. \cite{Lee2014}

\subsection{FFAFFURR} \label{Method-FFAFFURR}
 

Force field parameterization is an optimization problem with three challenging aspects: \cite{Aktulga2015}
\begin{enumerate}
\item The optimization problem has to be defined, which consists of the objective of the optimization and, following this, the selection of training data as well as force-field parameters to adjust.

\item In order to perform the force-field parameterization, a preferably automated procedure has to be implemented. The framework and algorithms used in FFAFFURR are explained in this paper. 
\item The obtained set of force-field parameters has to be validated against other data than the training data. 
\end{enumerate}

\noindent
Regarding item 1, the parameters of every energy term in a force-field have to be optimized since energy terms and parameters are interdependent and only adjusting a subset may cause inconsistency. 
Items 1 and 3, training and validation, heavily rely on high-quality data. We use DFT data for comparing potential energies and further validate by MD simulation.

Some practical points were considered when establishing the FFAFFURR framework: (i) the framework should be straightforward to set up and use, (ii) it should be easy to extend with other FF parameters or functional forms, and (iii) the result should be immediately usable by a molecular simulation package. 
FFAFFURR acts as a “wrapper” between the molecular mechanics package openMM \cite{Eastman2017} and the \textit{ab initio} molecular simulation package FHI-aims. \cite{BLUM20092175, HAVU20098367, Ren_2012} 
The code reads QM data directly from the output of FHI-aims and the output itself is a parameter file that can be processed by openMM. 
FFAFFURR is designed as the next step of the genetic algorithm package Fafoom. \cite{Supady2015} 
Conformers obtained by Fafoom through global search can be directly parsed to FFAFFURR. 
FFAFFURR is an open source tool and can be found at \url{https://github.com/XiaojuanHu/ffaffurr-dev/releases/tag/version1.0}.

\subsubsection*{Bond and angle parameterization}

$K_{ij}^{r}$, $K_{ij}^{\theta}$, $r_{ij}^0$ and $\theta_{ij}^0$ are empirical parameters of bond-stretching and angle-bending terms. 
The “spring” parameters $K_{ij}^{r}$ and $K_{ij}^{\theta}$ are unaltered in FFAFFURR.
The focus simply lies on the “torsional” and “non-bonded” parameters.
Bond-stretching and angle-bending terms intend to model small displacements away from the lowest energy structure. 
We adjust $r_{ij}^0$ and $\theta_{ij}^0$ by simply taking the average of the respective bond or angle over all local minima in the quantum chemistry data set.

\subsubsection*{Torsion angle parameterization}

The torsion angle term represents a combination of the bonded and nonbonded interactions. 
It has been reported that torsional parameters fitted to gas phase QM data perform similarly to those fitted to the experimental data. \cite{wang2017building} 
Although torsional parameters can be derived from vibrational analysis or using vibrational spectra as target data, this approach is complicated and requires a more elaborate treatment. \cite{pulay1979systematic, kumar2020ffparam, mayne2013rapid} 
In the case of the torsion term, force constants $V_n^{ij}$ and $V_{2imp}^{ij}$ can be tuned by LASSO or Ridge regression to minimize the difference between the FF and QM torsional energies. 
The ``torsions contribution'' from QM $\widetilde E^{\mathrm{QM}}_{\mathrm{torsions}}$ is calculated as:
\begin{equation}
\widetilde E^{\mathrm{QM}}_{\mathrm{torsions}} = E^{\mathrm{QM}}_{\mathrm{total}} - E^{\mathrm{FF}}_{\mathrm{nonbonded}}  - E^{\mathrm{FF}}_{\mathrm{bond}} - E^{\mathrm{FF}}_{\mathrm{angle}},
\end{equation}
where $E^{\mathrm{QM}}_{\mathrm{total}}$ represents the total energy of a conformer from a QM calculation, $E^{\mathrm{FF}}_{\mathrm{nonbonded}}$, $E^{\mathrm{FF}}_{\mathrm{bond}}$ and $E^{\mathrm{FF}}_{\mathrm{angle}}$ represent energies of nonbonded terms, bonded terms, and angle terms from FF calculations, respectively.

\subsubsection*{Electrostatic parameterization}\label{ele_param}

A key difference between FF parameter sets is the origin of the atomic partial charges. 
Deriving charges from QM data is widely used. 
The workflow of FFAFFURR tested three choices of partial charges: Hirshfeld, \cite{Hirshfeld1977, Bultinck2007} ESP \cite{Bultinck2007, Singh1984} and RESP \cite{Bayly1993} charges. 
The charge of each atom type of the force-field is defined as the average value of QM charges.
The scaling factor $f_{ij}$ used to scale the electrostatic interactions between the third neighbors (1,4-interactions) can also be adjusted by fitting to minimize the difference between the FF and QM energies.

\subsubsection*{LJ parameterization} \label{LJ_para}

Pair-specific Lennard-Jones (LJ) interaction parameters (referred to as NBFIX in the CHARMM force-fields) have been proven to better describe the interaction between cations and carbonyl groups of a protein backbone. \cite{Li2015} 
FFAFFURR employs pairwise Lennard-Jones (LJ) parameters instead of values determined by the combination rule.

In recent years, progress has been made in the calculation of pairwise dispersion interaction strength from the ground-state electron density of molecules. \cite{klimevs2012perspective, reilly2015van, cole2016biomolecular} 
The interatomic pairwise parameter $\sigma_{ij}$ can be derived using the atomic Hirshfeld partitioning scheme, which has already been used in the pairwise Tkatchenko-Scheffler vdW model. 
With the concept of the vdW radius, the LJ energy can be written as
\begin{equation} \label{eq_FF_vdw}
E_{\mathrm{vdw}} = \sum_{i<j} \varepsilon_{ij} \left [ \left ( \frac{R^{\mathrm{min}}_{ij}}{r_{ij}} \right)^{12} - 2\left (\frac{R^{\mathrm{min}}_{ij}}{r_{ij}} \right)^{6} \right ] f_{ij}, 
\end{equation}
where $R^{\mathrm{min}}_{ij}$ refers to the atomic distance where the vdW potential is at its minimum. 
With the definition of the free and effective atomic volume $V^{\mathrm{free}}$ and $V^{\mathrm{eff}}$, $R^{\mathrm{min}}_{ij}$ is estimated as the sum of effective atomic van der Waals radii of atom $i$ and atom $j$. 
The effective vdW radius of an atom is given by
\begin{equation}\label{R_eff}
R_{\mathrm{eff}}^0 = \left ( \frac{V^{\mathrm{eff}}}{V^{\mathrm{free}}} \right ) ^{1/3}R_{\mathrm{free}}^0,
\end{equation}
where $R_{\mathrm{free}}^0$ is the free-atom vdW radii that correspond to the electron density contour value determined for the noble gas on the same period using its vdW radius by Bondi. \cite{Bondi1964}
Pairwise $\sigma_{ij}$ can be calculated as
\begin{equation}
\sigma_{ij} = 2^{-1/6}R^{\mathrm{min}}_{ij}.
\end{equation}
The $\varepsilon_{ij}$ parameter from eq. \ref{eq_FF_vdw} can be tuned by fitting FF LJ energies to reproduce QM vdW energies by LASSO or Ridge regression.  

\subsubsection*{Deriving charge transfer parameters}

In all Zinc-finger proteins and most enzymes, Zn$^{2+}$ coordinates to four ligands.
However, due to the setup of the QM data set with monomeric and dimeric peptides, the cations have coordination numbers (CNs) of one or two. 
Therefore we added a correction factor for CN in eq. \ref{charge_transfer}
\begin{equation} \label{correction_ct}
\Delta q_{\mathrm{L-Me}}  = \frac{1}{\mathrm{CN}^k} ( a_L r_{\mathrm{Me-L}} + b_L ). 
\end{equation}
$k$, $a_L$, and $r^{\mathrm{cutoff}}$ can be adjusted by PSO. 
The target objective of fitting can be the QM potential energy, the QM interaction energy, or the electrostatic potential derived from electron densities. 
$b_L$ can be calculated with the assumption that charge transfer is zero at the cutoff distance.

\subsubsection*{Polarization energy}

To get the value of atomic polarizability $\alpha_i$ in eq. \ref{pol}, we use the definition of effective polarizability of an atom in a molecule, where the free-atom polarizabilitiy is scaled according to its close environment with a partitioning:
\begin{equation}
\alpha_{\mathrm{eff}} = \left (\frac{V^{\mathrm{eff}}}{V^{\mathrm{free}}} \right )\alpha^0_{\mathrm{free}},
\end{equation}
where $V^{\mathrm{eff}}$ and $V^{\mathrm{free}}$ are the same as in eq. \ref{R_eff}, and $\alpha^0_{\mathrm{free}}$ is the isotropic static polarizability. 
$\alpha_i$ is taken by averaging over all atoms with the same atom type in the quantum chemistry data set.
FFAFFURR also supports slightly adjusting $\alpha_i$ by fitting force-field energies to reproduce QM energies via PSO.

\subsubsection*{Boltzmann-type weighted fitting}

The quantum chemistry data set covers a wide range of relative energies.
By transitioning from, in our case, DFT to an additive force-field, even including charge transfer and polarization, we reduce dimensionality of the energy function and therewith the ability to correctly/fully represent the PES. Consequently, a force-field, describing, e.g., such a cation-protein system, cannot fully reproduce a DFT PES. Hence, it appears advisable to put focus on the accuracy of distinct areas of the PES.
RMSD between two surfaces is a common fitting criteria, but this approach gives more weight to areas of the energy surface with larger absolute values, while the real weight should more closely represent the Boltzmann weight of the energy surface. 
Consequently, we calculate Boltzmann-type weights and apply them as a scoring function.
The weighted RMSD, $wRMSD$, is given as:
\begin{equation}
wRMSD = \left [ \sum_{i=1}^{N} w_i ( E_i^{\mathrm{FF}} - \Delta E_i^{\mathrm{QM}}) ^2 \right ] ^{\frac{1}{2}},
\end{equation}
where RMSD is modified by including a Boltzmann-type factor,
\begin{equation}
w_i = A \exp \left [ \frac{-E_i^{\mathrm{QM}}}{\mathrm{RT}} \right],
\end{equation}
where A is the normalization constant (so that $\sum{w_i}$ = 1) and RT is a temperature factor that has no physical meaning in the context of this application, but affects the flatness of the distribution.
Our previous work \cite{Amin2020} has shown how Boltzmann-type weighted RMSD with an appropriate choice of RT can be utilized as the objective function for force-field parameter optimization. 
Therefore, we implemented Boltzmann-type weighted fitting in FFAFFURR by scaling the energies with the corresponding Boltzmann-type weights.

\subsection{Validation of new parameters}
\subsubsection*{Assessment of the energies} \label{Assess_energies}

To evaluate the performance of the parameterization, energies of conformers in the test set calculated with optimized parameters were compared to DFT energies by mean absolute errors (MAEs) and maximum errors (MEs). 
The MAE for the relative energies between FF energies and QM energies is calculated as
\begin{equation} 
\mathrm{MAE} = \frac{1}{N} \sum_{i=1}^{N} \arrowvert \Delta E_i^{\mathrm{FF}} - \Delta E_i^{\mathrm{QM}} + c \arrowvert \label{eq_MAE},
\end{equation}
where $N$ is the number of conformers in a given data set. 
$\Delta E_i$ refers to the energy difference between conformer $i$ and the lowest-energy conformer in the set. 
The adjustable parameter $c$ is used to shift the FF or QM energy hierarchies to one another to get the lowest MAE. 
ME is calculated as:
\begin{equation}
\mathrm{ME} =\max_{i \in N} \arrowvert \Delta E_i^{\mathrm{FF}} - \Delta E_i^{\mathrm{QM}} + c\arrowvert.
\end{equation}

\subsubsection*{Molecular dynamics simulations}

We performed MD simulations of the NMR structure 1ZNF \cite{lee1989three} with different parameter sets to evaluate the performance of FFAFFURR.
All MD simulations were performed using OpenMM7. \cite{Eastman2017} 
The structure of 1ZNF was placed in a cubic box of 68 {\AA} side length filled with TIP3P water. 
Four Cl$^{-}$ were added to neutralize the system. 
Then energy minimization was performed with the steepest descent minimization.
To equilibrate the solvent and ions around the protein, we continued 100 ps NVT and 100 ps NPT equilibration at a temperature of 300 K.
SHAKE constraints were applied to heavy atoms of the protein. 
Then independent MD simulations were performed with a time step of 2 fs. 
In all calculations, the long-range electrostatics beyond the cutoff of 12 {\AA} were treated with the Particle Mesh Ewald (PME) method. \cite{darden1993particle} The LJ cutoff was set to 12 {\AA}. 
The LJ and electrostatic interactions were computed every time step. 
For the simulations with the CTPOL model, charge transfer and induced dipoles were updated every 10 steps.
Covalent bonds and water angles were constrained.

\section{Results and discussion}

To assess the performance of FFAFFURR and describe which protocol to use to create a parameter set, we optimized the parameters of OPLS-AA with FFAFFURR and extended the OPLS-AA model by the CTPOL model. 
The quality of optimized parameters was assessed by examining the structural stability of the Zinc-finger motif in MD simulations.

\subsection{OPLS-AA parameterization}

Although studies have shown that it is difficult to implicitly incorporate the polarization effects into classical FFs, \cite{zhang2012modeling, Amin2020} fine-tuning parameters of fixed-charge models to describe cation-protein systems is still attractive due to its low computational cost and easier parameterization. 
Here we tested the performance of the fixed-charge model OPLS-AA parametrized using FFAFFURR. 
Five systems were tested: (1) AcAla$_{2}$NMe; (2) AcAla$_{2}$NMe+Na$^+$; (3) AcCys$^-$NMe; (4) AcCys$^-$NMe+Zn$^{2+}$; and (5) AcCys$^{-}_{2}$NMe+Zn$^{2+}$. 
AcAla$_{2}$NMe and AcAla$_{2}$NMe + Na$^+$ were used as reference models since the charge transfer and polarization effects caused by Na$^+$ are less than that of Zn$^{2+}$.
On the contrary, Cys$^{-}$ is one of the ligands that interact with Zn$^{2+}$ in proteins, and charge transfer between Cys$^{-}$ and Zn$^{2+}$ is significant.
For each system, 80 percent of the conformers were randomly selected as the training set, and the remaining 20 percent were used as the test set.

\begin{figure}
\vspace{0.0in}
\begin{center}
\includegraphics[width=1.0\textwidth]{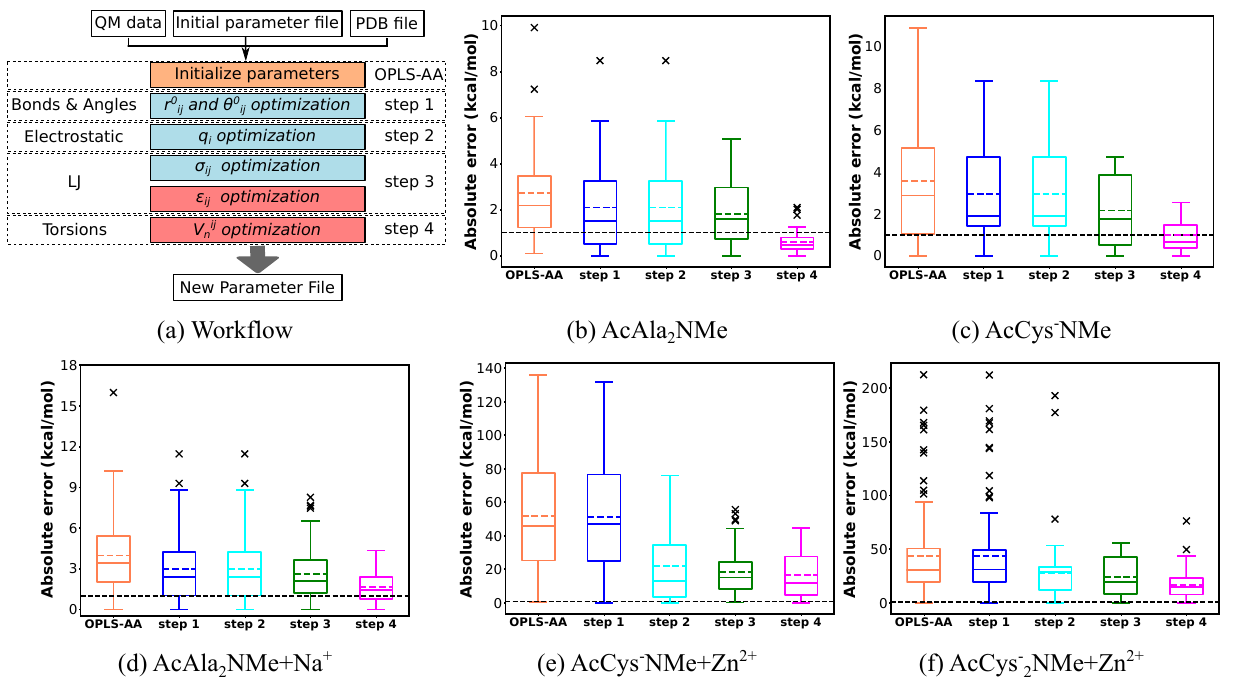}
\end{center}
\vspace{-0.2in}
\setlength{\belowcaptionskip}{0.2in}
\caption{(a) Workflow of the parameterization of OPLS-AA in four major steps. Different colors represent different fitting methods. Parameters in blue boxes are derived from DFT calculation, and parameters in red boxes are tuned by LASSO or Ridge regression. (b-f) Box plots of absolute errors of OPLS-AA parameterization major steps (OPLS-AA, step 1, step 2, step 3, step 4) for the test set of (b) AcAla$_2$NMe, (c) AcCys$^-$NMe, (d) AcAla$_2$NMe+Na$^+$, (e) AcCys$^-$NMe+Zn$^{2+}$ and (f) AcCys$^{-}_{2}$NMe+Zn$^{2+}$. The upper and lower lines of the rectangles mark the 75\% and 25\% percentiles of the distribution, the horizontal line in the box indicates the median (50 percentile), internal colored dashed line indicate the mean value, and the upper and lower lines of the “error bars” depict the 99\% and 1\% percentiles. The crosses represent the outliers. Black dashed line indicates the chemical accuracy, which is 1 kcal/mol. Note the large differences in scales in subfigures (b) to (f).
}
\label{full-opt-oplsaa}
\vspace{-0.2in}
\end{figure}

We first demonstrate the functionality of FFAFFURR on the example of OPLS-AA parameterization. The key steps of OPLS-AA parameterization are briefly described in Figure \ref{full-opt-oplsaa} (a). 
We showed the ability to reproduce PES by optimizing parameters of bonds, angles, electrostatic interactions, LJ interactions, and torsional interactions. 
Users can choose which energy items to adjust according to their needs. 
In Figure \ref{full-opt-oplsaa} (a), the parameters in blue boxes are derived from DFT calculations and the parameters in red boxes are fitted by LASSO or Ridge regression as described in Sections \ref{Method-optimization} and \ref{Method-FFAFFURR}. 
Here, we only tested RESP partial charges, the LASSO method in deriving $\varepsilon_{ij}$, and Ridge regression in deriving $V_{n}^{ij}$. 
The parameters derived from DFT calculations are tuned first because they are considered fixed with respect to changes of the other parameters, then different tuning orders of the parameters for the other energy terms in the FF formula are tested to choose the order that gives the smallest errors between DFT and FF energies. The final order of the protocol is shown in Figure \ref{full-opt-oplsaa} (a).

Figure \ref{full-opt-oplsaa} (b-f) shows the comparison of FF energies with optimized parameters after each step in Figure \ref{full-opt-oplsaa} (a).
Noticeably, charges for AcAla$_2$NMe, AcCys$^-$NMe and AcAla$_2$NMe+Na$^+$ were not altered since the original charges yielded errors lower than average RESP charges from QM calculations, while average RESP charges were employed for AcCys$^-$NMe+Zn$^{2+}$ and AcCys$^{-}_{2}$NMe+Zn$^{2+}$. 
Figure \ref{full-opt-oplsaa} (e) and (f) indicate that using average RESP charges significantly reduces absolute errors for AcCys$^-$NMe+Zn$^{2+}$ and AcCys$^{-}_{2}$NMe+Zn$^{2+}$. 
This could be due to the capture of charge transfer to some extent. 
In the case of AcAla$_2$NMe and AcCys$^-$NMe, the MAEs were improved from \mbox{2.72 kcal/mol} and 3.59 kcal/mol to 0.61 kcal/mol and 0.98 kcal/mol, respectively, which is well within the 1 kcal/mol chemical accuracy. 
In the case of AcAla$_2$NMe+Na$^+$, the MAE was improved from 3.99 kcal/mol to 1.67 kcal/mol. 
Although the optimized MAE is above the chemical accuracy, the maximum error is significantly reduced. 
However, in the cases of AcCys$^-$NMe+Zn$^{2+}$ and AcCys$^{-}_{2}$NMe+Zn$^{2+}$, the MAEs were improved from 51.75 kcal/mol and 43.47 kcal/mol to 16.8 kcal/mol and 16.59 kcal/mol, respectively. 
Although these are by numbers great improvements, the MAEs are much higher than for the other systems. Calculations based on parameters of such quality have no predictive power.
This confirms the necessity of explicitly including charge transfer and polarization effects to describe divalent ion-dipeptide systems.
We note that for dipeptides and dipeptides with monovalent cation systems, optimization of torsional parameters has the greatest impact on the improvement of the accuracy.
Previous studies by some of us \cite{baldauf2013cations, ropo2016trends} have shown that mono- and divalent cations strongly modify the preferences of torsion angles. 
While for dipeptides with divalent cations, apparently, the adjustment and treatment of charge interactions plays the most important role. 
This further confirms that the capture of charge transfer and polarization is crucial for the accurate description of systems with divalent cation. 
We also note that the maximum errors are greatly reduced after the parameterization of LJ interactions of the five systems.

\subsection{CTPOL parameterization}

\begin{figure}
\vspace{0.0in}
\begin{center}
\includegraphics[width=1\textwidth]{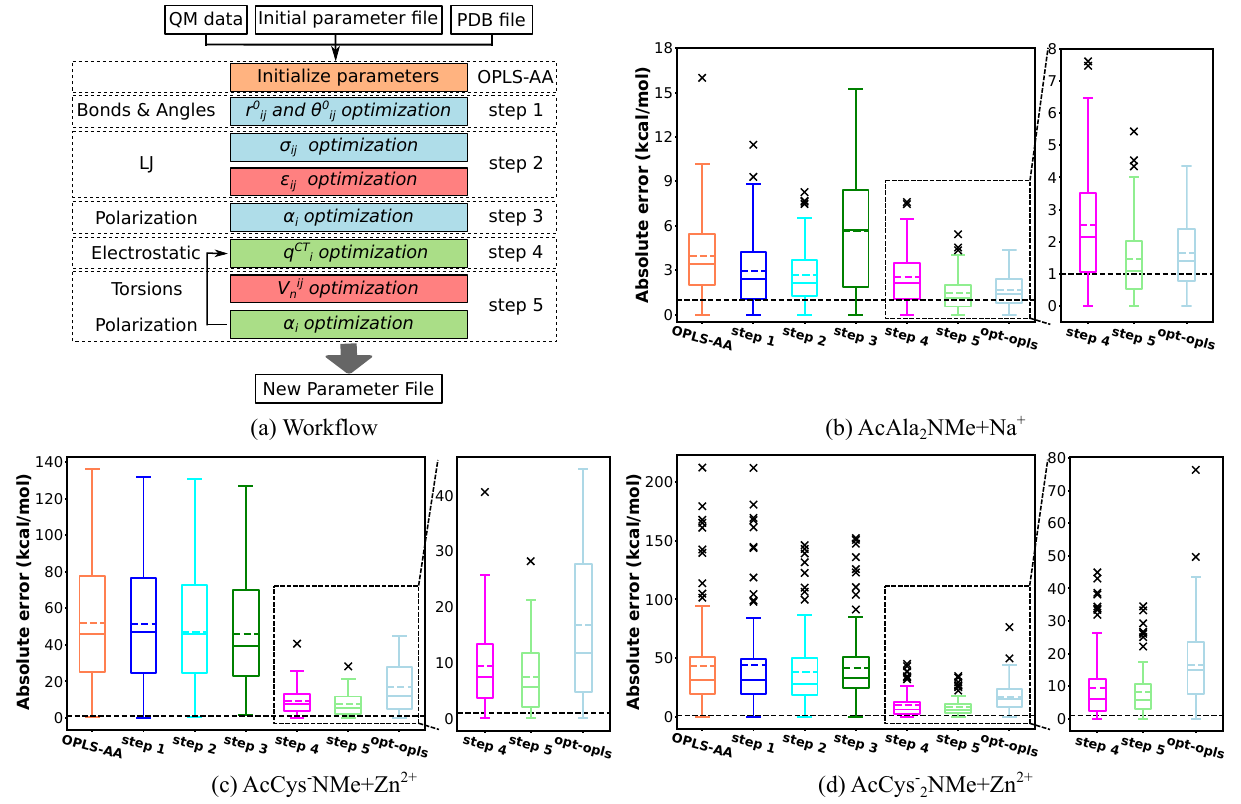}
\end{center}
\vspace{-0.2in}
\setlength{\belowcaptionskip}{0.2in}
\caption{ (a) Workflow of full CTPOL parameterization in five major steps. Different colors represent different fitting methods. Parameters in blue boxes are derived from DFT calculation, parameters in red boxes are tuned by LASSO or Ridge regression, and parameters in green boxes are tuned by PSO. (b-d) Box plots of absolute errors of CTPOL parameterization major steps (OPLS-AA, step 1, step 2, step 3, step 4, step 5) and OPLS-AA with full optimized parameters (opt-opls) for test set of (b) AcAla$_2$NMe+Na$^+$, (c) AcCys$^-$NMe+Zn$^{2+}$ and (d) AcCys$^{-}_{2}$NMe+Zn$^{2+}$. The upper and lower lines of the rectangles mark the 75\% and 25\% percentiles of the distribution, the horizontal line in the box indicates the median (50 percentile), internal colored dashed line indicate the mean value, and the upper and lower lines of the “error bars” depict the 99\% and 1\% percentiles. The crosses represent the outliers. Black dashed line indicates the chemical accuracy, which is 1 kcal/mol.}
\label{full-opt-ctpol}
\vspace{-0.2in}
\end{figure}

The CTPOL model introduces both local polarization and charge-transfer effects into classical force-fields. 
We investigated the performance of the CTPOL model on the cation-dipeptide systems: AcAla$_2$NMe+Na$^+$, and two challenging systems AcCys$^-$NMe+Zn$^{2+}$ and AcCys$^{-}_{2}$NMe +Zn$^{2+}$. 
The major steps of the CTPOL parameterization workflow are depicted in Figure \ref{full-opt-ctpol} (a). 
Following the methodology of OPLS-AA optimization, parameters unaffected by others are adjusted first, then different orders are tested to choose the order that gives smallest errors between DFT and FF energies. Charges are taken from OPLS-AA in step 1 to step 3. In step 4, charge transfer was introduced.
As already mentioned, the parameters in blue boxes are derived from DFT calculations and the parameters in red boxes are fitted by LASSO or Ridge regression. 
The parameters in green boxes are obtained by PSO. 
Noticeably, $\alpha_i$ is tuned twice. 
In step 3, $\alpha_i$ is taken as the average effective polarizability calculated from the \textit{ab initio} method. 
In step 5, we tried to slightly tune $\alpha_i$ by PSO.
An additional round of parameterization from step 4 to step 5 can be performed to better optimize the FF parameters.

Absolute errors of each step in Figure \ref{full-opt-ctpol} (a) are illustrated in Figures \ref{full-opt-ctpol} (b-d). 
Absolute errors of optimized OPLS-AA (opt-opls) are also shown in Figure \ref{full-opt-ctpol} to compare the performance of FFAFFURR on OPLS-AA and CTPOL models. 
As shown in Figure \ref{full-opt-ctpol}, the introduction of polarization effects in step 3 didn't improve the accuracy much, and the errors of the AcAla$_2$NMe+Na$^+$ system even increased. 
This may be due to the fact that classical force-fields already take some account of polarization effects, since the charges come from fitting to reproduce quantum mechanical or experimental electrostatic field distributions. \cite{sakharov2009force} 
Including charge transfer from ligand atoms to the cation reduces atomic charges, therefore compensating for the electrostatic potential. 
Not surprisingly, errors are significantly reduced after including charge transfer as displayed in Figure \ref{full-opt-ctpol}. 
After the parameterization, the MAEs of AcAla$_2$NMe+Na$^+$, AcCys$^-$NMe+Zn$^{2+}$ and AcCys$^{-}_{2}$NMe+Zn$^{2+}$ reached 1.45 kcal/mol, 7.42 kcal/mol, and 8.12 kcal/mol, respectively. 
In contrast, the MAEs of the optimized OPLS-AA are 1.67 kcal/mol, 16.8 kcal/mol, and 16.59 kcal/mol, respectively. 
Apparently, the inclusion of charge transfer and polarization effects better describes systems involving cations than classical force-fields, especially for systems with divalent cations.

\begin{figure}[htbp]
\vspace{0.0in}
\begin{center}
\includegraphics[width=0.75\textwidth]{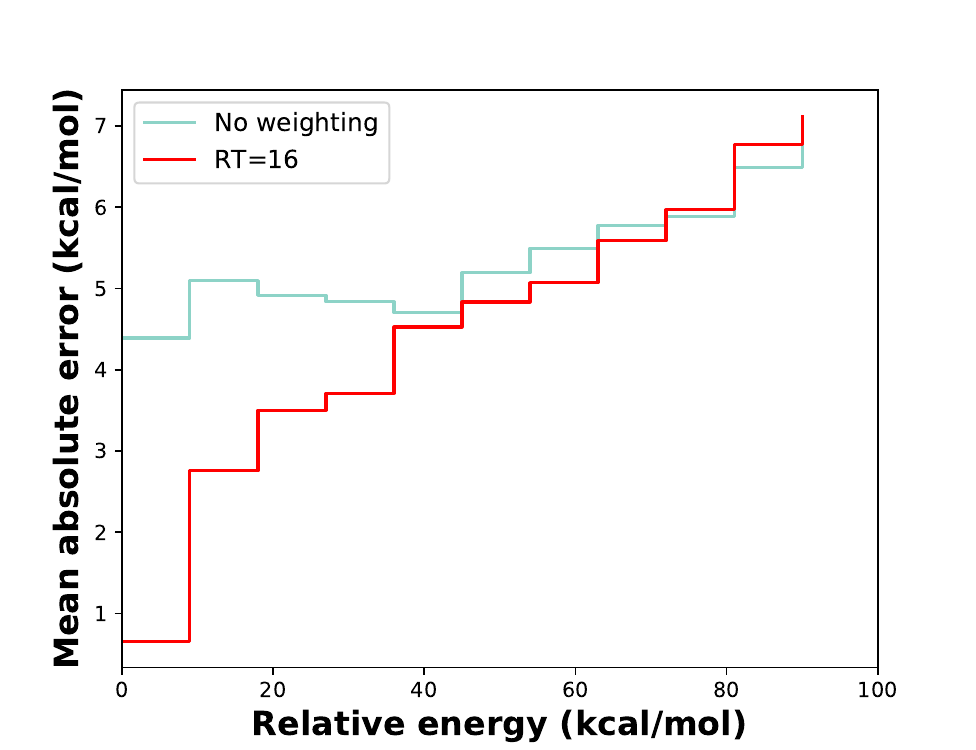}
\end{center}
\vspace{-0.2in}
\setlength{\belowcaptionskip}{0.2in}
\caption{Absolute errors of optimized FF energies with respect to QM energies by weighted/unweighted fitting of AcCys$^-$NMe+Zn$^{2+}$ system. The height of the bar represents the mean absolute error for conformers whose relative energies are smaller than the right node of the bar.}
\label{BW_Cys-Zn}
\vspace{-0.2in}
\end{figure}

To focus the fitting on the low-energy part of the PES, we applied Boltzmann-type weights to the scoring function during the fitting of the charge transfer parameters. 
In Figure \ref{BW_Cys-Zn}, the AcCys$^-$NMe+Zn$^{2+}$ system is taken as an example. 
Figure \ref{SI_BW_Cys-Zn} shows the Boltzmann-type weights ($w_i$) along QM relative energies with different temperature factor (RT) values. 
For all RT, the weight decreases as the relative energy increases, 
but increasing RT decreases the weights on low-energy conformations. 
Figure \ref{BW_Cys-Zn} shows the difference in mean absolute errors between unweighted fitting and weighted fitting with RT = 16. 
In Figure \ref{BW_Cys-Zn}, the height of the bar represents the mean absolute error for conformers whose relative energies are smaller than the right node of the bar.
Interestingly, the weighted fitting improves accuracy substantially in the low-energy region, while high-energy regions do not get worse.

\FloatBarrier

\subsection{Validation with molecular dynamics simulations}




The 1ZNF PDB structure \cite{lee1989three} is one of the first Zinc-finger structures to be resolved experimentally. 
It is also the simplest, containing only 25 amino acids and one \ch{Cys2His2} Zn$^{2+}$ binding domain  where the Zinc ion is in a stable coordination geometry consisting of cysteine sulfurs and histidine nitrogens in the first coordination shell (see Figure \ref{fig:MD-rmsd}).
Due to its compact size, 
the 1ZNF structure provides an ideal case study for an MD validation of a FFAFFURR parameterization workflow.
One potential application of FFAFFURR to this system is to optimize selected parameters for the interaction center (Figure \ref{fig:MD-rmsd} bottom-left), since that is the region of most complexity.


In this paper, we used an approach similar to Li \emph{et al.}, \cite{li2016mcpb} giving the residues in the interaction center unique residue names to distinguish them from similar residues in the rest of the protein. 
This allows us to target only atom types within the binding domain for parameterization, without affecting the parameters of similar atom types away from the binding site.

Four parameter sets were tested with MD in this study, as described in Table \ref{tbl:parameter_sets}. For the unparameterized OPLS-AA force-field, we observed unbinding of the two histidine residues from the Zn$^{2+}$ interaction center, as shown in Figure \ref{structure-MD}, almost immediately after the start of the simulation.
To try and prevent this, we optimized pair-wise LJ parameters between atoms in HisD and Zn$^{2+}$. 
The parameters that are optimized are listed in Table \ref{tbl:opt_LJ_params_full}. The LJ parameters between atoms in Cys and Zn$^{2+}$ are kept untouched since we haven't seen strange behaviors between Cys and Zn$^{2+}$.
The optimized LJ parameters were used in opt-OPLS-AA and opt-CTPOL sets. In the CTPOL and opt-CTPOL models, charge transfer was introduced for S/N/O/Zn atoms in the binding site, and polarization effects between non-hydrogen atoms and Zn$^{2+}$ were added. Oxygen is included as some of the structures in the ab-initio dataset have Zn$^{2+}$ interacting with a peptide oxygen atom. 

\begin{table}
  \caption{Parameter sets used for MD simulation.
  The determination of LJ parameters from FFAFFURR is described in \ref{LJ_para}. 
    optimized parameters are listed in Table \ref{tbl:opt_LJ_params_full} and \ref{tbl:CTPOL_params}.}
  \label{tbl:parameter_sets}
  \begin{tabular}{lll}
    \hline
      Parameter set& Pair-wise LJ parameters of atoms in HisD and Zn$^{2+}$  &  CT + POL \\
    \hline
    OPLS-AA & original  &  No \\
    opt-OPLS-AA& from FFAFFURR  & No \\
    CTPOL  & same as OPLS-AA & Yes \\
    opt-CTPOL & from FFAFFURR  & Yes \\
    \hline
  \end{tabular}
\end{table}

\subsubsection*{Backbone structure and binding domain are better preserved with CTPOL}

\begin{figure}[htbp]
\includegraphics[width=\textwidth]{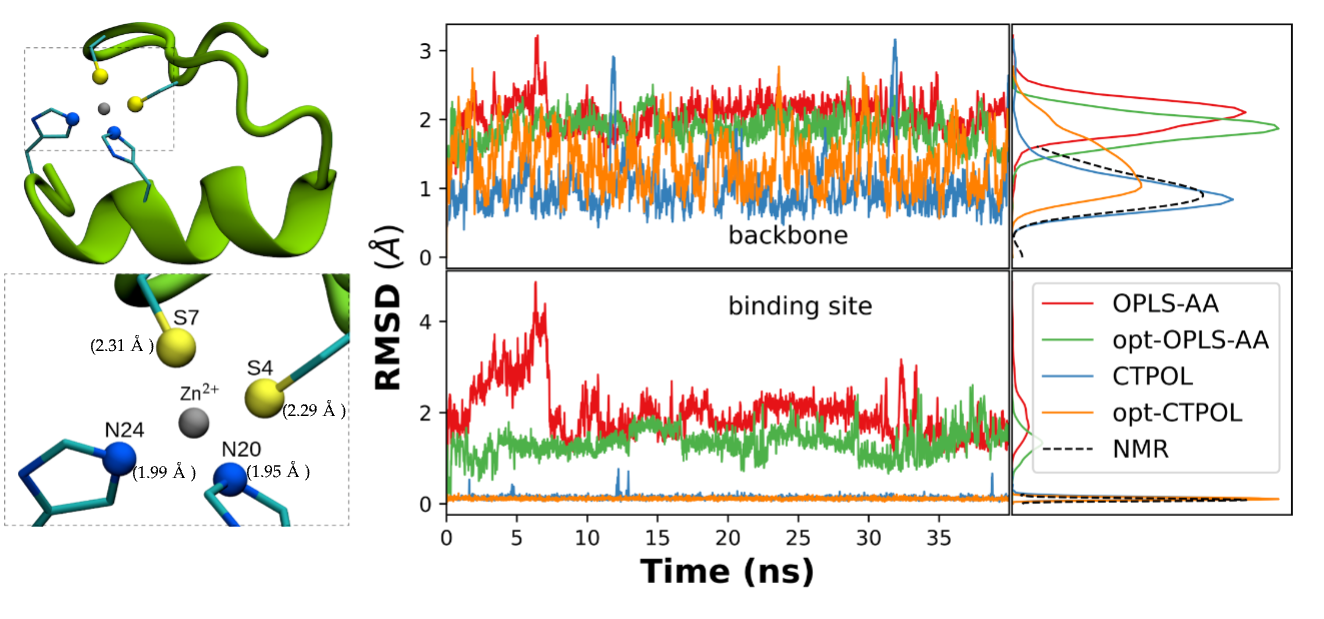}
\caption{\textit{Top-Left}: The protein structure of 1ZNF, with the backbone represented by a ribbon, and the Zn\ch{^{2+}}-binding site shown explicitly. \textit{Bottom left}: Zoom in of the site, with distances of the coordinating atoms relative to Zn\ch{^{2+}}. 
The sulfurs are from Cys4 and Cys7, while the nitrogens are the NE2 nitrogens of His20 and His24.
\textit{Right}: RMSDs of MD trajectories from the NMR structure of 1ZNF (Model 1), calculated for different parameter sets, for backone (top) and interaction site (bottom).  
The densities of RMSD values are shown on the right, using Kernel Density Approximation \cite{parzen1962kde, rosenblatt1956kde}, where the dashed line is the RMSD distribution obtained from NMR data of 1ZNF with respect to the first model of the PDB.}
\label{fig:MD-rmsd}
\end{figure}

We ran three 40 ns long simulations with each of the four models listed in Table \ref{tbl:parameter_sets}.
We also used the 37 experimental NMR structures of 1ZNF to compare structural features between our simulations and NMR observations. 
Figure \ref{fig:MD-rmsd} shows the RMSD of each of the parameter sets, using the first model of the NMR structures as a reference. 
In the same figure, we also plot the RMSD of the 37 NMR models with respect to the same first model to see how much variation occurs among those.

It is clear from Figure \ref{fig:MD-rmsd} that both, the overall structure and binding domain, are in better agreement with the NMR structures when charge transfer and polarizability are taken into account. 
With opt-OPLS-AA, there is a marginal but noticeable improvement over OPLS-AA, but in both OPLS-AA and opt-OPLS-AA force-fields the binding domain breaks apart. This is evident from the RMSD of the backbone, as shown in the bottom panel of Figure \ref{fig:MD-rmsd}. This is primarily due to the Histidines breaking away from the binding with Zn$^{2+}$, as supported by Figure \ref{SI:zn-n_v_time}.


\begin{figure}[htbp]
\centering
\subfigure[OPLS-AA]{
    \includegraphics[width=0.22\textwidth]{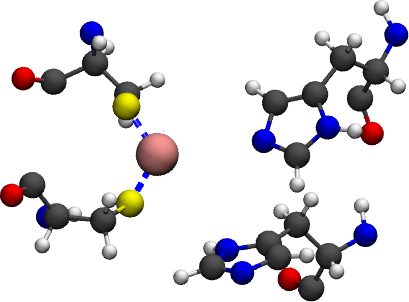}
}
\subfigure[opt-OPLS-AA]{
    \includegraphics[width=0.22\textwidth]{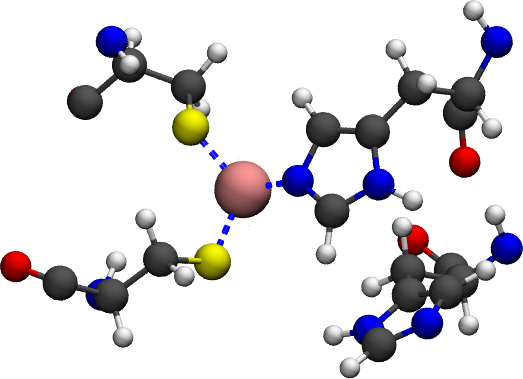}
}
\subfigure[CTPOL]{
    \includegraphics[width=0.22\textwidth]{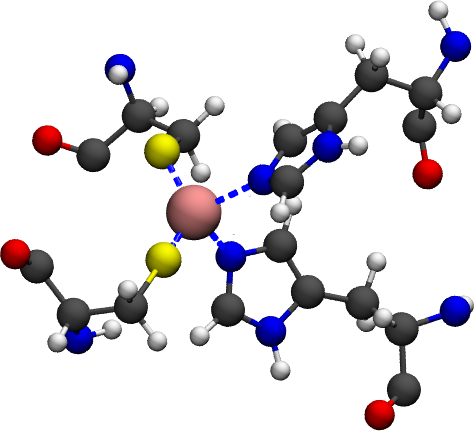}
}
\subfigure[opt-CTPOL]{
    \includegraphics[width=0.22\textwidth]{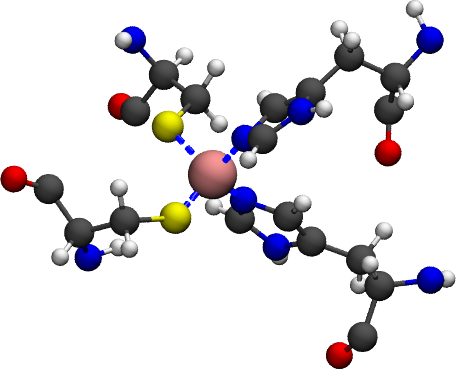}
}
\caption{Snapshots showing the conformation of binding site after 40 ns of simulation.}
\label{structure-MD}
\end{figure}

The RMSDs of OPLS-AA and opt-OPLS-AA deviate far from the NMR model, particularly the RMSDs of the binding site only.
We observed in our simulations that with OPLS-AA, the two histidine residues in the binding site stray uncharacteristically far from \ch{Zn^{2+}}. 
Even  with optimization of the pair-wise LJ parameters of Zn$^{2+}$ and histidine (opt-OPLS-AA), we observed one of the histidines escaping the binding domain. Figure \ref{structure-MD} (a) and (b) shows snapshots of such conformations after 40 ns.
Similar problems with binding domain stability have been observed in previous studies, where the Zn$^{2+}$ escapes from the coordination center in non-polarizable FF simulations. \cite{zhang2012modeling, donini2000calculation}

However, both CTPOL and opt-CTPOL preserve the binding domain of \ch{Zn^{2+}}, with both histidines and both cysteines coordinating the \ch{Zn^{2+}} ion throughout the 40 ns simulations (snapshots of Figure \ref{structure-MD} (c) and (d)).
This emphasizes that explicitly including charge transfer and polarization effects is critical for a proper description of the binding domain, and hence the overall structure of Zinc-fingers.

\FloatBarrier

\subsubsection*{LJ parameterization makes the CTPOL model more robust}

\begin{figure}
    \centering
    \includegraphics[width=0.8\textwidth]{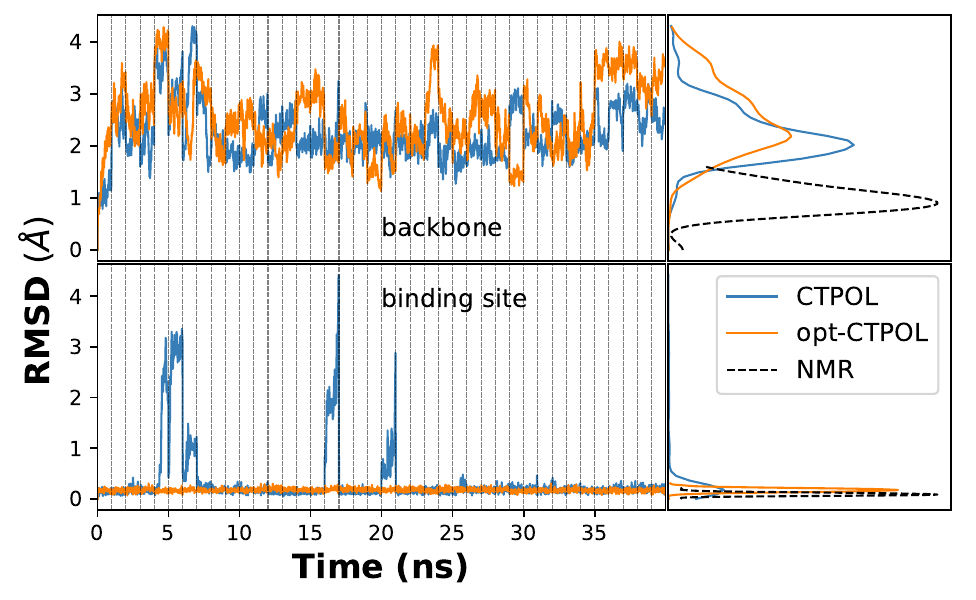}
    \caption{RMSD of CTPOL and opt-CTPOL vs 1st model of NMR, with 40 trajectories of 1 ns concatenated into one.
    The dotted lines represent concatenation boundaries of the trajectories.}
    \label{fig:ctpol_rmsd}
\end{figure}

To evaluate the effect of optimized pair-wise LJ parameters we compared the CTPOL model without any LJ parameterization (CTPOL) to the CTPOL model with LJ parameterization (opt-CTPOL). 
From Figure \ref{fig:MD-rmsd}, it may appear that such optimization has little effect, and in fact may slightly worsen the overall structure due to the higher RMSD of the backbone. However,
while both models preserve the interaction center much better than OPLS-AA and opt-OPLS-AA,
opt-CTPOL appears to produce a much more stable binding domain than CTPOL. This can be seen when we recompute RMSD after varying the initial conditions. 
To test the impact of initial conditions, we ran 40 independent 1ns long simulations, with the initial frame randomly chosen from a 4 ns MD simulation and random initial velocities. These are reasonable initial conditions that should exhibit similar behavior, as they are taken from a simulation.  
Figure \ref{fig:ctpol_rmsd} shows that the 40 ns trajectory of CTPOL using the NMR structure as the starting point is more or less stable.
However, when running simulations from different initial conditions, this stability is not guaranteed, as seen from the spikes in RMSD. 
On the other hand, opt-CTPOL appears to be stable for all initial conditions.

\begin{figure}
    \centering
    \includegraphics[width=0.8\textwidth]{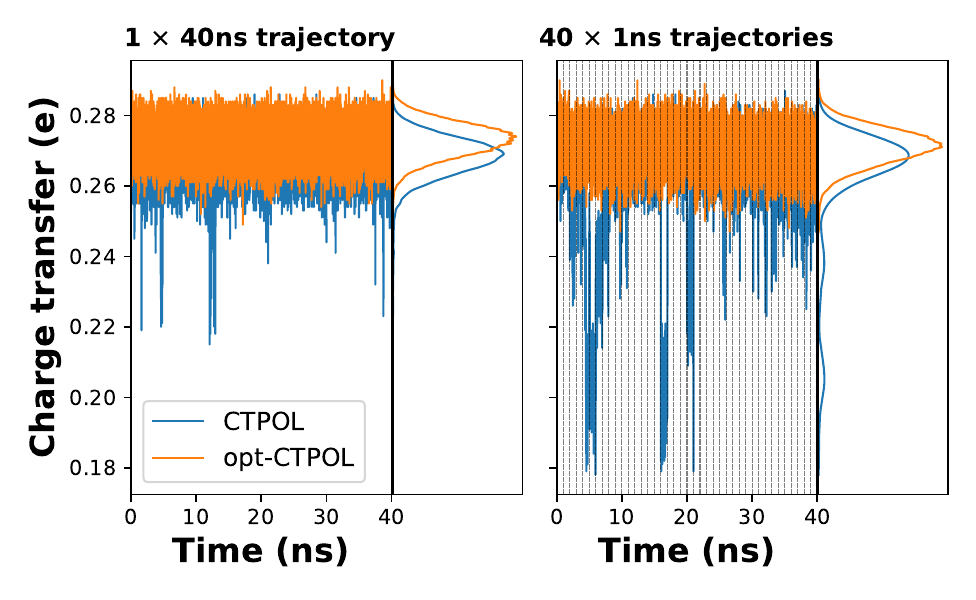}
    \caption{Charge transfer as a function of time for (left) a continuous 40 ns trajectory from one stable initial structure, 
    and (right) 40 independent 1 ns simulations concatenated together. The dashed vertical lines mark the concatenation boundaries.
    The $40 \times 1$ ns simulations were started from different initial conditions randomly chosen from a continuous MD simulation, with randomized velocities. }
    \label{fig:ct_vs_time}
\end{figure}

\begin{figure}
    \centering
    \textbf{Charge transfer and relevant distances in CTPOL}\par\medskip
    \includegraphics[width=0.8\textwidth]{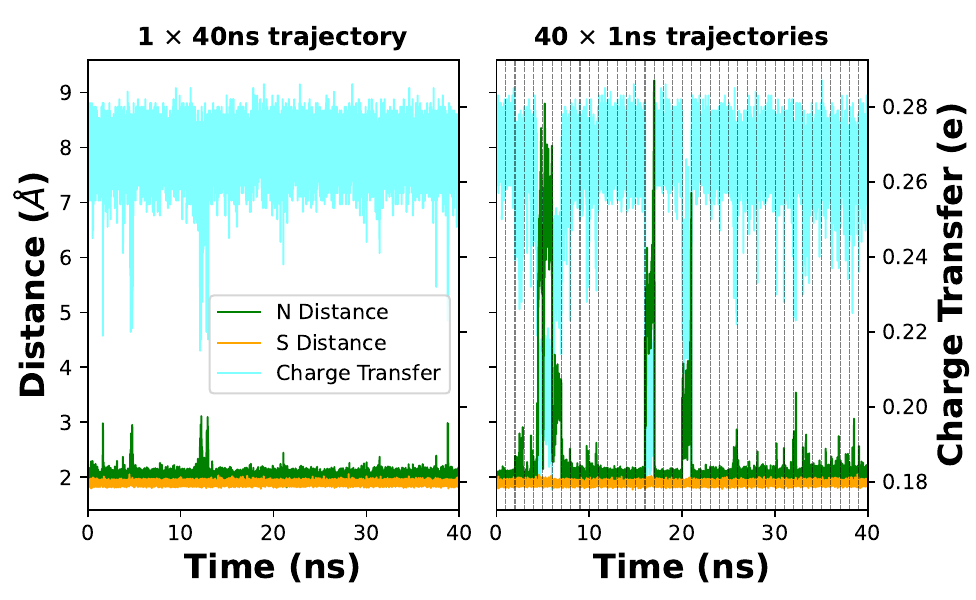}
    \caption{Coordinating nitrogen and sulfur distances (left y-axis) and charge transfer (right y-axis) vs time for a continuous trajectory (left) and 40 independent concatenated trajectories.
    In cyan, we have the charge transfer, in green, the average of the distances of Zn-N20 and Zn-N24, and in yellow the average of the distances of Zn-S4 and Zn-S7. Out of the 40 independent simulations, the average distance of Zn-N20/24 rises above 3 {\AA} 8 times. 
    }
    \label{fig:ct_dists_v_time}
\end{figure}

A reason for this is the abnormal charge transfers to Zn\ch{^{2+}} in CTPOL as seen in Figure \ref{fig:ct_vs_time}. This occurs around the same time as the binding domain fluctuations in Figure \ref{fig:ctpol_rmsd}. 
A closer inspection of the distances between Zn\ch{^{2+}} and coordinating nitrogens (Figure \ref{fig:ct_dists_v_time}) reveals that these fluctuations are perfectly correlated with these distances. 
As the binding site breaks down, the coordinating histidines containing these nitrogens move far away, as much as 9 {\AA} away, but the sulfurs remain in close proximity at all times. 
At such distances, the charge transfer contribution of the nitrogens drops to zero, and the only contribution is from the sulfurs, and hence the lower total charge transfer.
However, opt-CTPOL appears to have no such fluctuation in either the 40 ns or 40 $\times$ 1 ns trajectories. 

These unfolding events within 1ns occur about 20\% of the time for CTPOL, thus making CTPOL without LJ-optimization unreliable. 

\FloatBarrier
\subsubsection*{opt-CTPOL shows improvement with a caveat to be addressed in the future}

\begin{figure}
    \centering
    \textbf{Continuous 40 ns trajectory}
    \subfigure[Radial Distribution Functions]{
    \includegraphics[width=0.8\textwidth]{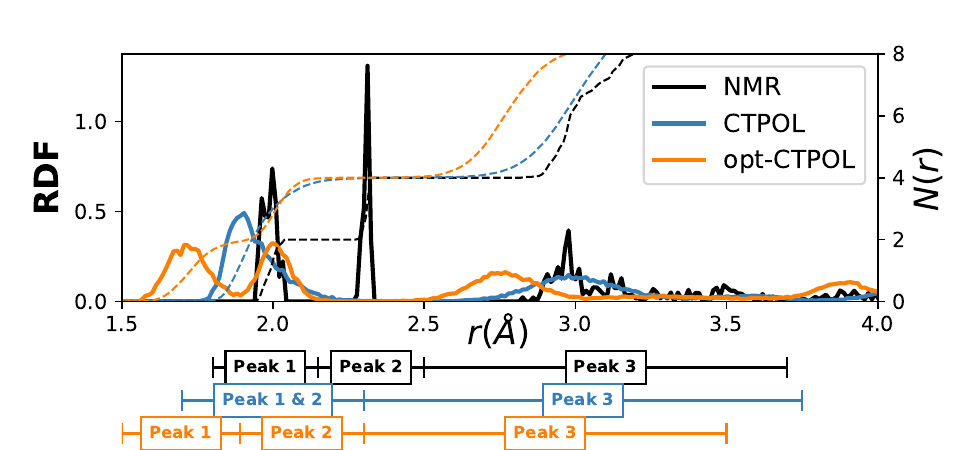}
    }\\
    \subfigure[Peak Breakdown]{
    \includegraphics[width=0.8\textwidth]{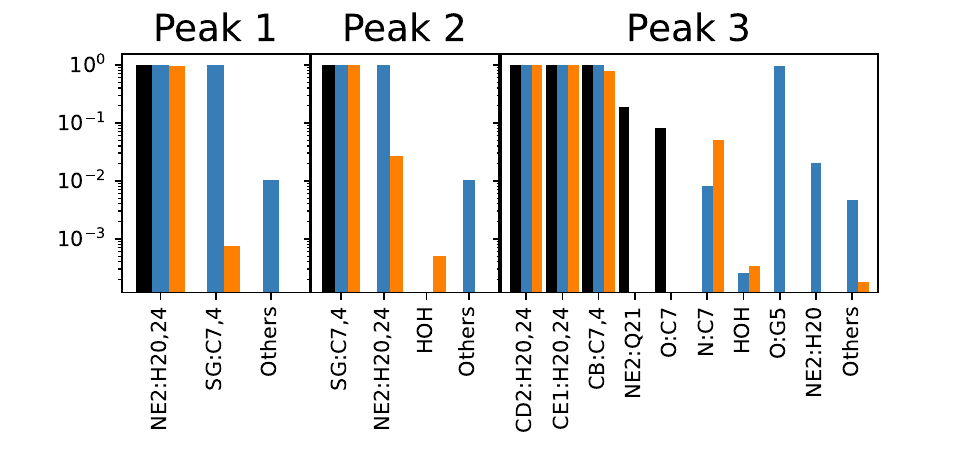}
    }
    \caption{\textbf{Coordination analyses of continuous 40 ns trajectory.} a) RDF and coordination number ($N(r)$) of all non-hydrogen protein atoms, with the distance ranges of selected peaks. The solid lines are the RDF (left $y$ axis), and the dashed lines are the corresponding $N(r)$ (right $y$ axis). Composition of each peak, where atoms of the same type and residue are lumped  together.
    The $y$ axis represents the average fraction of conformations in which each of the atoms appears within the peak range.
    }
    \label{fig:rdf_peaks}
\end{figure}

\begin{figure}
    \centering
    \textbf{$40\times 1$ns trajectory}
    \subfigure[Radial Distribution Functions]{
    \includegraphics[width=0.8\textwidth]{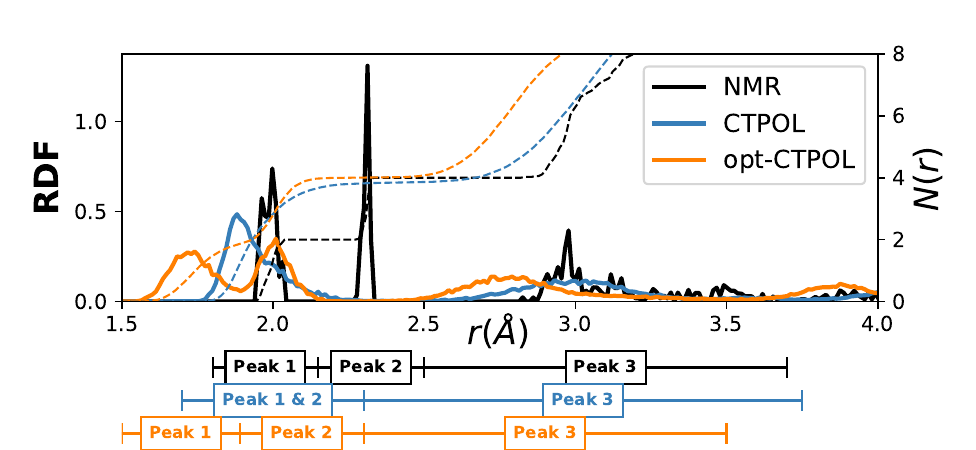}
    }\\
    \subfigure[Peak Breakdown]{
    \includegraphics[width=0.8\textwidth]{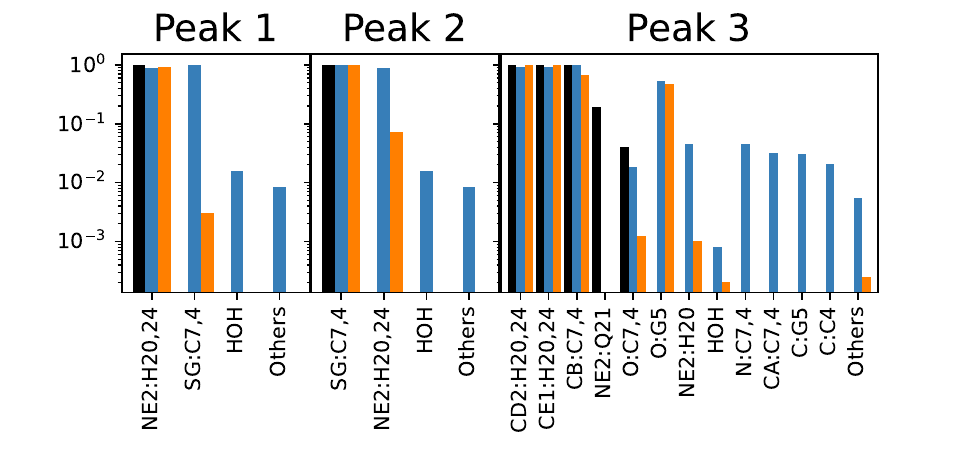}
    }
    \caption{\textbf{Coordination analyses of $40\times1$ ns trajectory.} a) RDF and coordination number ($N(r)$) of all non-hydrogen protein atoms, with the distance ranges of selected peaks. Solid lines are RDF, and dashed lines are the $N(r)$. b) Composition of each peak, where atoms of the same type and residue are lumped  together. 
    The $y$ axis represents the average fraction of conformations in which each of the atoms appears within the peak range.
    }
    \label{fig:rdf_peaks_40_1}
\end{figure}


To evaluate how parameters affect the coordination of \ch{Zn^{2+}}, we plotted the radial distribution function of non-hydrogen protein atoms around the cation in Figure \ref{fig:rdf_peaks} (top).
We can see immediately that NMR and opt-CTPOL have a similar peak structure, but the distances are shorter in opt-CTPOL. 
In CTPOL, the first and second peaks, containing Nitrogens and Sulfurs respectively, overlap completely and are indistinguishable.
In the NMR models, the first and second peaks at 2.0 and 2.3 $\AA$ correspond to Zn$^{2+}$-N(His) and Zn$^{2+}$-S(Cys), respectively. 
In contrast to CTPOL, the opt-CTPOL peaks are distinct, with only a small percentage (< 2\%) of trajectories showing Nitrogens in the 2nd peak dominated by Sulfur. 
These features are also seen in similar analyses of the $40\times 1$ ns trajectories (Figure \ref{fig:rdf_peaks_40_1}). 
Based on the analyses in this manuscript we can conclude that CTPOL does not reproduce NMR binding domain as well as opt-CTPOL even for the stable continuous 40 ns trajectory.

After identifying the peaks, and selecting a range of distances (Figure \ref{fig:rdf_peaks} (top)), we determined which atoms comprise each peak and what fraction of the trajectory these atoms remain in that peak, as shown in Figure \ref{fig:rdf_peaks} (bottom). 
The 1st and 2nd peaks in CTPOL appear to be contaminated by other atom types which do not appear in NMR peaks at all. 
In the 40 $\times$ 1 ns trajectory, since CTPOL binding site has been shown to break apart in a few cases, it is no surprise that water also appears in Peak 1 of CTPOL (Figure \ref{fig:rdf_peaks_40_1} (bottom)).
The opt-CTPOL model has no other atom types in the first peak, and only relatively few others in the 2nd peak not present in NMR.

We should note that the NMR model we used does not contain any explicit water molecules. 
To determine if water could be present in the binding site, we looked at 15 Zinc-finger X-ray crystallography structures from the Protein Data Bank\cite{berman2000pdb} (PDB) website (\href{http://www.rcsb.org/pdb/}{http://www.rcsb.org/pdb/}) to find binding sites which are similar to this one (see \ref{SI:tbl:other_pdbs} for a full list). 
We looked at binding sites which had a total of 2 histidines and 2 cysteines, similar to 1ZNF.  
We found 8 binding sites from the 15 crystal structures, and the smallest water distance to \ch{Zn^{2+}} was 4.38 {\AA}, well outside even the 3rd peak range in the NMR models. 
We further relaxed the matching criterion for the binding site to any binding site that contains a total of 4 histidines or cysteines (i.e., the number of coordinating histidines and cysteines sum to 4, but does not have to be 2 each).
This resulted in a total of 60 binding sites.
From these, we found the smallest water distance to be 3.98 {\AA}, still beyond the peak 3 range.

Thus, the inclusion of water in the 1st and 2nd peaks, as is the case in CTPOL model, is uncharacteristic of Zn-finger binding sites of similar nature to 1ZNF. 
The opt-CTPOL model does a better job of keeping the water outside these peaks, with only a small fraction of water in the 2nd peak. 

\FloatBarrier
\subsubsection*{Angle and distance distributions}
\begin{figure}
  \begin{minipage}[c]{0.49\textwidth}
    \includegraphics[width=\textwidth]{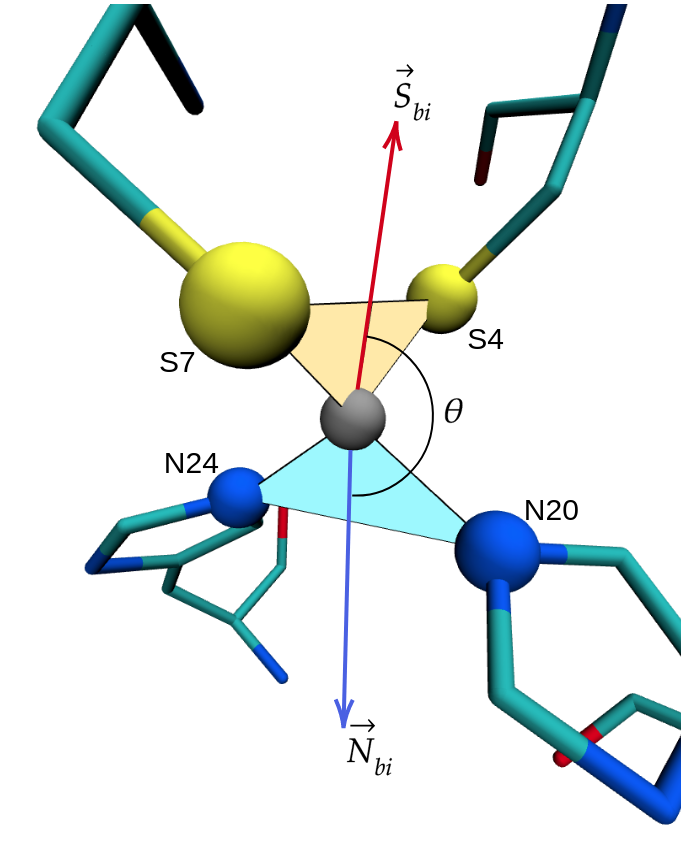}
  \end{minipage}\hfill
\begin{minipage}[c]{0.49\textwidth}
    \caption{
    Binding site with \ch{Zn^{2+}} at the center (grey atom), the sulfurs from Cys7 (S7) and Cys4 (S4), the NE nitrogens from His20 (N20) and His24 (24). 
    Hydrogens have been removed for clarity. 
    The yellow triangle on top has vertices on Zn, S4 and S7, while the blue triangle at the bottom has vertices on Zn, N20, and N24. 
    The angle between the planes of these triangles are used for plotting the distributions in Figure \ref{fig:dihedral_bisector}.
    The red and blue arrows ($\vec{S}_{bi}$ and $\vec{N}_{bi}$) are vectors that bisect angles S7-Zn-S4 and N20-Zn-N24 respectively.
    The distributions of angle $\theta$ between these two bisectors are plotted on Figure \ref{fig:dihedral_bisector} (b).
    The distributions of some of the distances between the 5 atoms shown in this figure are shown on Figure \ref{fig:distances}, while the distributions for some of the angles are shown in Figure \ref{fig:angles}.
    } \label{fig:binding_angles}
  \end{minipage}
\end{figure}

To further evaluate the stability and accuracy of the binding domain in the CTPOL and opt-CTPOL frameworks, we analyzed a number of geometric quantities which are defined in Figure \ref{fig:binding_angles} and its caption. 
Here we only consider the 40 ns continuous trajectory for which the binding domain is stable for CTPOL, since these geometric quantities would not make sense for the $40\times1$ns trajectory where the binding domain destabilizes.

\begin{figure}
    \centering
    \includegraphics[width=.7\linewidth]{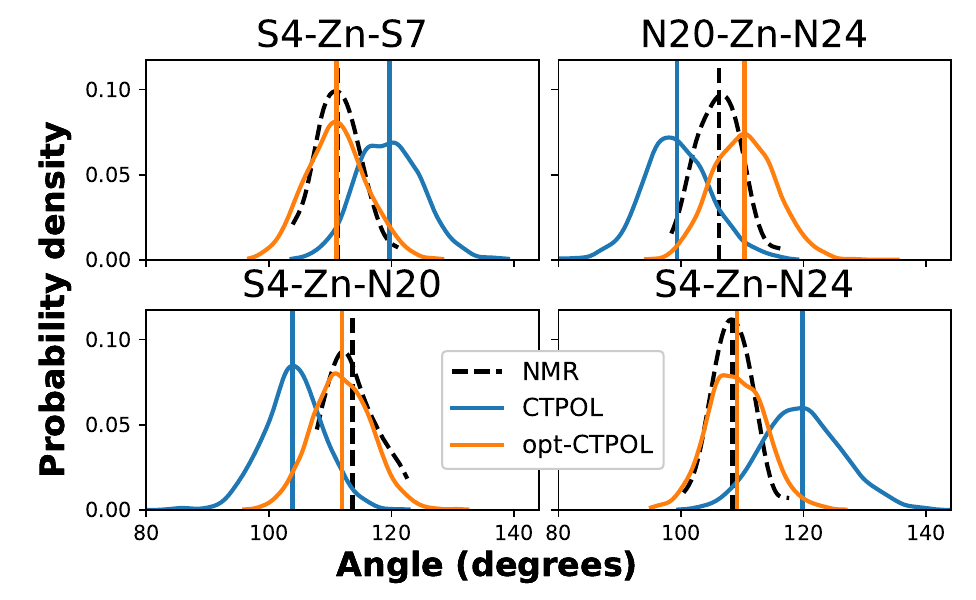}
    \caption{Probability distribution of angles over the continuous 40 ns trajectories of CTPOL (blue) and opt-CTPOL (orange) and over 37 NMR models (black dashed). 
    The corresponding atoms are depicted in Figure \ref{fig:binding_angles}. The distributions were calculated using Kernel Density Estimation \cite{parzen1962kde, rosenblatt1956kde}. The vertical lines represent the averages of each distribution. 
    }
    \label{fig:angles}
\end{figure}

\begin{figure}
    \centering
\subfigure[Dihedral distribution]{
      \includegraphics[width=.35\linewidth]{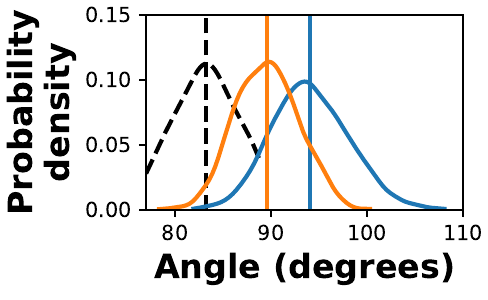}
}%
\subfigure[Bisector angle distribution]{
      \includegraphics[width=.35\linewidth]{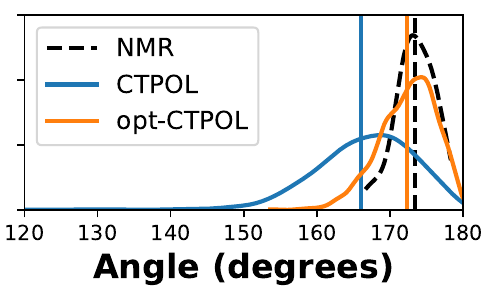}
}%
\caption{(a) Probability distributions of angle between the S7-Zn-S4 and N24-Zn-N20 planes as depicted in Figure \ref{fig:binding_angles}. (b) Angle between S4-Zn-S7 and N20-Zn-N24 bisectors, which are depicted in Figure \ref{fig:binding_angles} as $\vec{S}_{bi}$ and $\vec{N}_{bi}$, respectively.
}
\label{fig:dihedral_bisector}
\end{figure}

Figure \ref{fig:angles} shows the distribution of most of the angles that the coordinating atoms make with Zn$^{2+}$.
Additionally, Figure \ref{fig:dihedral_bisector} (a) shows the distributions of angles between the planes shown in Figure \ref{fig:binding_angles},
and Figure \ref{fig:dihedral_bisector} (b) shows the distributions of the angles between the bisectors, also defined in Figure \ref{fig:binding_angles}.
It is quite clear that opt-CTPOL reproduces the NMR distributions of angles as well or better than CTPOL.
The distribution of the S4-Zn-S7 angle appears to agree particularly well with NMR, as does the angle between the bisectors. 
While the CTPOL 40 ns trajectory showed a slightly better overall RMSD from Figure \ref{fig:MD-rmsd}, it is clearly not reproducing these angles as well as opt-CTPOL.
This implies that opt-CTPOL is maintaining the shape of the binding domain better, which is in accordance with the RDF distribution and peak analysis of Figure \ref{fig:rdf_peaks}.

\begin{figure}
    \centering
    \includegraphics[width=.8\linewidth]{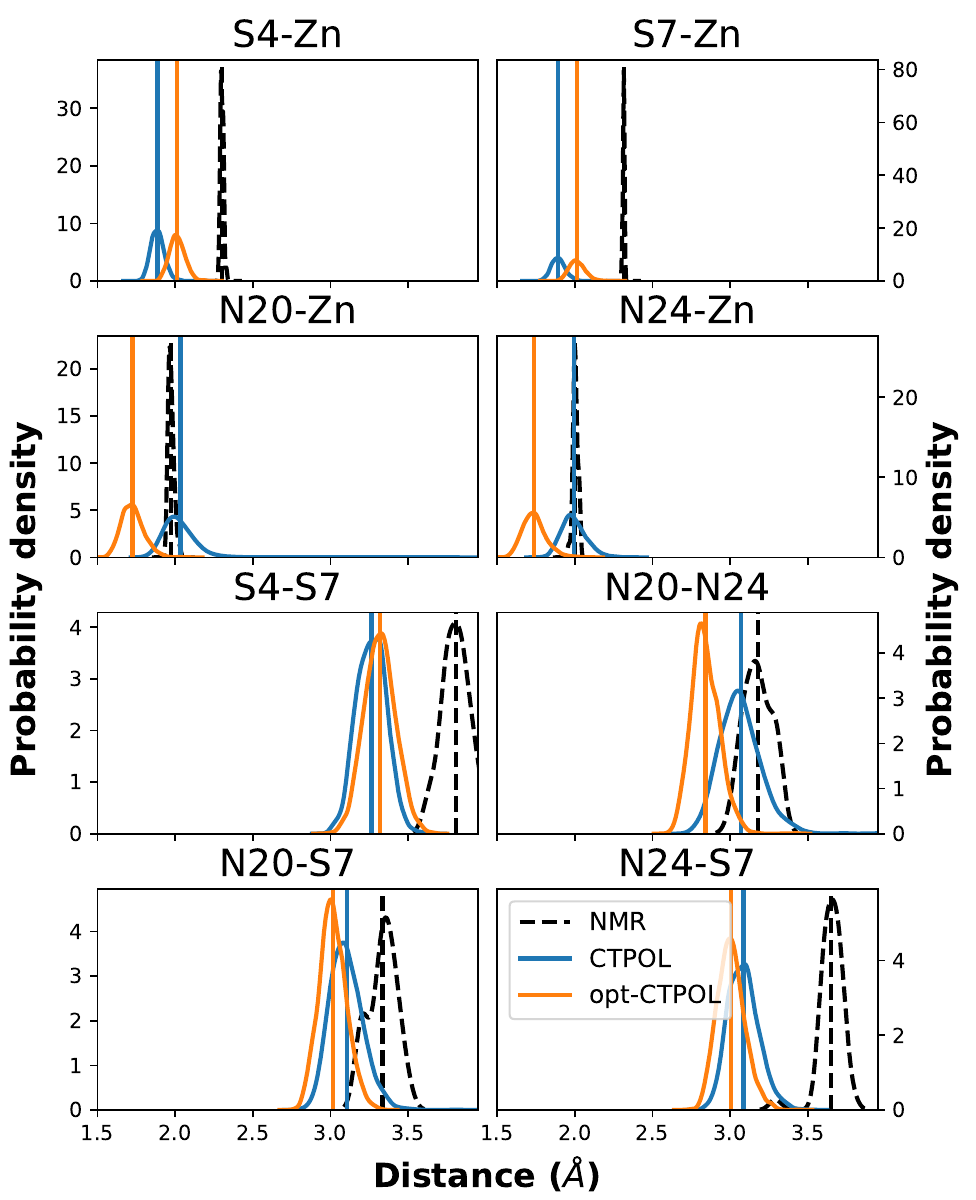}
    \caption{Probability distribution of distances (using Kernel Density Estimation \cite{parzen1962kde, rosenblatt1956kde}) over entire trajectory (for simulations) and over 37 models (for NMR data).}
    \label{fig:distances}
\end{figure}

Furthermore, we see from Figure \ref{fig:distances} that the distances of the opt-CTPOL binding domain are consistently shorter than those of the experimental NMR structures. 
This is in line with the RDF analysis of Figure \ref{fig:rdf_peaks}, where we see a similar peak structure of opt-CTPOL, but at shorter distances. 
On the other hand, CTPOL distances do not appear to have a consistent relation to the NMR distances. 
For instance, the distances of S*-Zn and N*-Zn (top left) show that opt-CTPOL distances trend the same way as NMR, i.e., the N*-Zn distances are significantly shorter than S*-Zn distances.
For CTPOL, it turns out to be almost the opposite, with plenty of overlap between the two distributions, and thus their 1st and 2nd peaks in Figure \ref{fig:rdf_peaks} also overlap. 

\FloatBarrier

\subsubsection*{Issues we observe and their possible origins}

\begin{figure}
    \centering
    \includegraphics{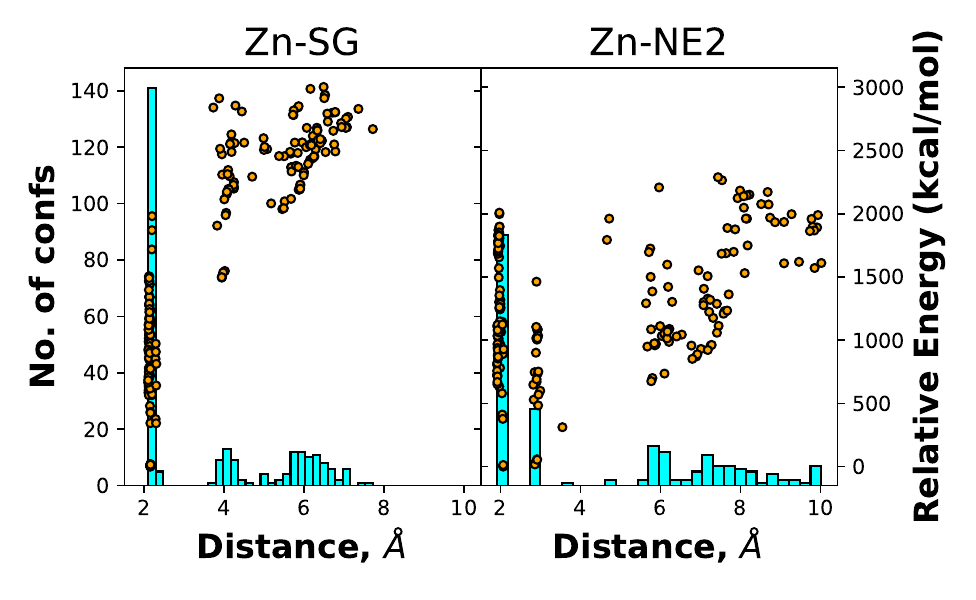}
    \caption{Distribution of \textit{ab-initio} distances, with bars representing histograms of the numbers of conformers (left axis) over distance range, and points representing the distance and energy of individual conformers (right axis).
    The plot on the left takes distances between Zn$^{2+}$ and S(Cys) from AcCys$^-$-NMe + Zn$^{2+}$ \textit{ab-initio} conformations. 
    The plot on the right takes distances between Zn$^{2+}$ and N(His) from AcCys$^-$-His + Zn$^{2+}$ \textit{ab-initio} conformations.}
    \label{fig:qm-distances}
\end{figure}

The main issue we face is clearly the contraction of distances in the Zn$^{2+}$-binding site in comparison to NMR data.
To investigate the source of this issue, we first analyzed the relevant distances in the \textit{ab initio} data set, as depicted in Figure \ref{fig:qm-distances}. 
We see that the Zn-N distances are mostly around the 2 {\AA} mark, and Zn-S distances are mostly around 2.3 \AA, as it is indicated by the maxima of the histograms.
In both cases, they cover a wide range of energies as well, suggesting that these are preferred distances for various conformations of the dipeptide.
However, these are (certainly almost exclusively) minima on the PES.

The larger distances in Figure \ref{fig:qm-distances} are due to coordination of Zn with other atoms (such as carbonyl oxygen). 
To make a meaningful comparison with the 1ZNF system, we further filtered the QM data down to where the nearest atom to Zn is atom SG or atom NE2, since that is the case in the 1ZNF system. 
From the filtered data, we computed the average distances, and compare them in Table \ref{tab:avg-distances}, where it is clear that the QM data agrees well with NMR, unlike opt-CTPOL, which systematically gives shorter distances.


\begin{table}[htb]
\begin{tabular}{r|cccc}

\textbf{Pair} & \textbf{QM} & \textbf{NMR} & \textbf{CTPOL} & \textbf{opt-CTPOL} \\ \hline
\textbf{N-Z}  & 1.95   & 1.99     & 2.01       & 1.73            \\ 
\textbf{S-Z}  & 2.28   & 2.31     & 1.89       & 2.01           \\ 
\end{tabular}
\caption{\label{tab:avg-distances}Average distance data.}
\end{table}


Given that the shorter distances are not an artefact that was carried from the gas phase QM data, in the following we discuss a few likely causes for this behavior of our model:
\begin{itemize}
    \item  We demonstrate in this paper (see Figure \ref{full-opt-ctpol}) that our approach can properly match energy hierarchies. However, the focus on equilibrium geometries may result in insufficient treatment / parameterization of repulsive (off-equilibrium) geometries. The effect may even be amplified due to the Boltzmann-weighting used in the parameter optimization process.

    \item The Lennard-Jones-(12,6)-potential is intended to model pairwise  repulsive and non-Coulomb attractive interactions among atoms. However, due to the nature of classical force-field formulations, it may also contain aspects of polarization, dipole-dipole interactions etc. These aspects may be incorporated in parameters as well as in the functional form (e.g. in the 12-6 form).

    \item Although we are matching MM energies at the exact same conformations, we do not know if the MM minima themselves correspond to those conformations. It is possible that we are simply matching the MM energies at the QM minima, but the MM minima may be in different states altogether, in this case, those that result in shorter ligand distances.
    \item None of the \textit{ab initio} references contain both Sulfur and Nitrogen in the coordination of Zn$^{2+}$, or even specifically S-N, N-N, and S-S pairwise interactions. We only parameterize Zn$^{2+}$-ligand LJ interactions in this work as an example, but it is possible that optimizing ligand-ligand interactions will improve the result.

\end{itemize}


\FloatBarrier
\section{Conclusion and outlook}
The availability of sufficiently accurate force-field parameters for cation-peptide systems is a major obstacle in metalloprotein simulations. 
One approach to facilitate the development of new force-field parameters is to construct tools to derive parameters from QM calculations.
The benefits of such an approach was shown in previous work \cite{Amin2020} on the QM-driven parameterization of Drude and CTPOL models  for ion-protein interactions in MD simulations.

Since the explicit Drude model includes polarization as a degree of freedom subject to forces, it requires shorter time steps and a dual thermostat and it introduces additional parameters such as the mass of Drude particles and spring constants.
CTPOL -- as an implicit model --  requires the minimization of dipole moments at every time step, but allows for normal time steps.
Both models exhibit  their own challenges and -- in particular our re-implementation of CTPOL -- their own room for improvements.


FFAFFURR is developed as a python tool to facilitate the parameterization of classical and polarizable CTPOL models. 
In this paper, we chose to parameterize OPLS-AA as an example, although the tool can be adjusted to work with other similar force-fields such as CHARMM and AMBER once the code is generalized. 
It automatically parses QM calculation outputs from FHI-aims and generates parameter files that can be directly processed by the molecular dynamics package OpenMM.
FFAFFURR also allows users to choose which energy terms to adjust.

We utilized an extensive data set of model peptide-cation conformers from DFT calculations. 
Structures cover an extended relative energy range and all required properties for the parameterization were extracted.  
Due to the sheer size of the utilized data set -- by means of number of conformers and energy range -- we think that we have covered sufficient individual structural diversity to properly derive parameters for the FF energy terms.
The performance of optimized parameters in each energy term was evaluated by comparing of FF energies and QM potential energies. 

We showed that the CTPOL model outperforms OPLS-AA in terms of the accuracy of reproducing the QM energy hierarchies for divalent-dipeptide systems. 

One potential usage of FFAFFURR is the rapid construction of FFs for troublesome metal centers in metalloproteins. 
We tested this function by performing MD simulations on the 1ZNF Zinc-finger protein\cite{lee1989three} and comparing simulation results with NMR models. 
With the parameters optimized from FFAFFURR, we found that CTPOL much better reproduces the overall structure of the protein, while with OPLS and opt-OPLS, the Zn$^{2+}$ binding site unravels. 
However, to better stabilize and reproduce structural features of the binding domain, LJ optimization (opt-CTPOL) was necessary, since CTPOL alone had some shortcomings in correctly reproducing the binding domain, or keeping it stable under various initial conditions.
The LJ optimization resulted in coordination composition and geometry that better agrees with the NMR models than CTPOL alone.

On the other hand, the optimization of LJ does lead to a shrunken binding domain. While we briefly discussed reasons for this behavior above, we focus here on possible remedies:
\begin{itemize}
    \item A possible lack of off-equilibrium geometries in the parameterization could be tested first by omitting the Boltzmann-weighting. This would give more impact to higher-energy structures that are more likely to feature at least partly more compressed structures. Maybe this already would lead to improved geometries.

    \item The repulsive part of the Lennard-Jones potential could be better captured by locally exploring the PES by adding points through dislocating atoms in the cation interaction site. This could either be realized by carefully altering Cartesian coordinates by small values in the $+$/$-$ space directions or through short DFT-based MD simulations. This would also capture whether or not the MM-minima are at those conformations.
    
    \item The functional form of the Van der Waals potential may be better generalized by alternatives to the Lennard-Jones form which allow for better tuning of the shape of the repulsive components. For example, the Mie potential is a generalized version of Lennard-Jones where the exponents and coefficients of each term can be varied. The Buckingham potential is another alternative where the repulsive term is replaced with an exponential decay with variable amplitude and decay rate. Such adjustments of the Van der Waals potential also requires a data base that contains sufficient off-equilibrium geometries.
    \item Lastly, the dataset could be expanded to contain models with more than one amino-acid side chain, in order to capture  complex coordination structures, allowing us to parameterize other pairwise interaction such as Sulfur and Nitrogen, and not just Zn$^{2+}$ and ligand atoms.

\end{itemize}


\noindent
In summary, FFAFFURR has a wide range of functions to facilitate and perform parameter optimization in peptide systems. 
It can provide better additive force-field parameters for systems with no cations, as seen from Figures \ref{full-opt-oplsaa}b and \ref{full-opt-oplsaa}c. 
Additionally, FFAFFURR and can provide almost all the functions required for the cation-peptide parameterization process, including for CTPOL force-fields. 
FFAFFURR helps users remove labor-intensive steps in force-field optimization. 

Despite the success of FFAFFURR in this study, we see several directions to discuss in future research. 
Note that only the parameters of the Zinc-finger protein interaction center were optimized with FFAFFURR in the MD simulation, while the standard OPLS-AA parameters were used for the rest of the protein. 
While our study indicates the compatibility of the optimized parameters with the standard FF parameters, this may have to be investigated in more detail in a future study. 
One characteristic of FFAFFURR is that it can be employed to derive parameters for a specific system. 
This helps to grasp the specific environment of the system. 
However, QM calculations are required when a new system is under investigation. 
We created a data set of cation-dipeptides containing several divalent cations, which can be automatically parsed to FFAFFURR. \cite{hu2022better} 
If the user's system goes beyond the scope of the dataset, an in-house genetic algorithm package Fafoom \cite{Supady2015} can be used to generate conformers and do the QM generation fast and automatically.

\begin{acknowledgement}

The authors thank:
\begin{description}
    \item [The China Scholarship Council] for providing X.H. with a doctoral fellowship;
    \item[The Federal Ministry of Education and Research of Germany] for providing funding for the project STREAM (“Semantische Repräsentation, Vernetzung und 333 Kuratierung von qualitätsgesicherten Materialdaten”, ID: 16QK11C); and
    \item[MITACS] for the MITACS Globalink Research Award which funded the visit of K.S.A. to the lab of X.H. and C.B.
    
\end{description}

\end{acknowledgement}

\section*{Available data and code}

The reference data can be found on the NOMAD repository via the DOI: \url{https://doi.org/10.17172/NOMAD/2023.02.03-1}.\cite{Hu2023NOMAD}
FFAFFURR can be found at: \url{https://github.com/XiaojuanHu/ffaffurr-dev/releases/tag/version1.0}.
Code for implementing CTPOL in OpenMM: \url{https://github.com/XiaojuanHu/CTPOL_MD}. \cite{Hu2023CTPOLMD}




\providecommand{\latin}[1]{#1}
\makeatletter
\providecommand{\doi}
  {\begingroup\let\do\@makeother\dospecials
  \catcode`\{=1 \catcode`\}=2 \doi@aux}
\providecommand{\doi@aux}[1]{\endgroup\texttt{#1}}
\makeatother
\providecommand*\mcitethebibliography{\thebibliography}
\csname @ifundefined\endcsname{endmcitethebibliography}
  {\let\endmcitethebibliography\endthebibliography}{}

\newpage\null\thispagestyle{empty}
\begin{suppinfo}

\renewcommand\thefigure{S\arabic{figure}}

\setcounter{figure}{0}

\begin{figure}[htbp]
\vspace{0.0in}
\begin{center}
\includegraphics[width=0.75\textwidth]{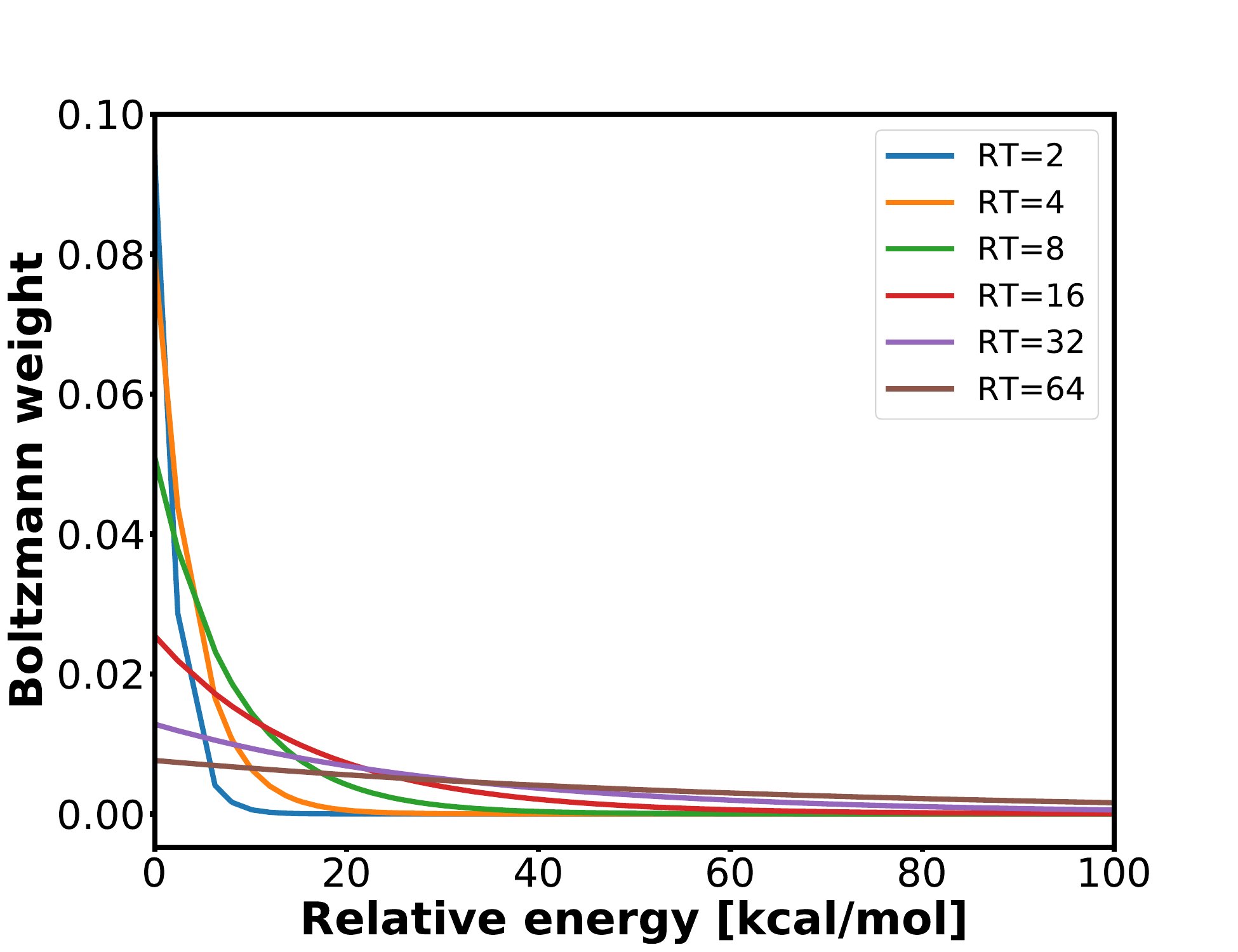}
\end{center}
\vspace{-0.2in}
\setlength{\belowcaptionskip}{0.2in}
\caption{Boltzmann-type weights vs. relative QM energies at various RTs of AcCys$^-$NMe+Zn$^{2+}$ system.}
\label{SI_BW_Cys-Zn}
\vspace{-0.2in}
\end{figure}

\renewcommand\thetable{S\arabic{table}}
\setcounter{table}{0}

\begin{table}
  \caption{Atom types in HisD+Zn$^{2+}$ and Cys$^{-}$+Zn$^{2+}$. }
  \label{tbl:atom_type_hisD}
  \begin{tabular}{lr | lr}
    \hline
    HisD+Zn$^{2+}$ & & Cys$^{-}$+Zn$^{2+}$ &  \\
    \hline
    Atom  & Atom type & Atom  & Atom type \\
    \hline
    C   & 2177  & C & 1177  \\
    CA & 2166  &  CA & 1166\\
    CB  & 2446 & CB & 1148 \\
    CD2 & 2448 & H & 1183\\
    CE1   & 2447  & HA & 1086 \\
    CG & 2449  & HB2 & 1085 \\
    H  & 2183  & HB3 & 1085 \\
    HA & 2086 & N & 1180 \\
    HB2   & 2085 & O & 1178  \\
    HB3 & 2085 & SG & 1142 \\
    HD1  & 2445 \\
    HD2 & 2091 \\
    HE1   & 2092   \\
    N & 2180  \\
    ND1  & 2444  \\
    NE2 & 2452 \\
    O &  2178 \\
    Zn & 834 \\
    \hline
  \end{tabular}
\end{table}

\begin{table}
  \caption{ The optimized LJ parameters. Epsilon = 0 means the LJ interaction is neglected. LASSO tends to focus on only important factors while neglecting insignificant ones. }
  \label{tbl:opt_LJ_params_full}
  \begin{tabular}{lrlr}
    \hline
    Type1  & Type2 & Sigma (nm)  & Epsilon (kJ/mol) \\
    \hline
    2178   & 834  & 0.31933 & 0.00024413  \\
    2448 & 834  &  0.331094 & 0 \\
    2183  & 834 & 0.32642 & 0.001277 \\
    2446 & 834 & 0.32934 & 0.03138\\
    2177   & 834  & 0.330867 & 0 \\
    2092 & 834  & 0.288564 & 0 \\
    2091 & 834  & 0.288564 & 0 \\
    2180  & 834  & 0.319954 & 0 \\
    2447 & 834 & 0.329767 & 0 \\
    2444   & 834 & 0.31992 & 0.25885  \\
    2445 & 834 & 0.29663 & 6.7250 \\
    2085  & 834 & 0.294726 & 0 \\
    2086  & 834 & 0.294726 & 0 \\
    2184 & 834 & 0.325209 & 0 \\
    2452 & 834 & 0.32598 & 0.00153   \\
    2166 & 834 & 0.331252 & 0.01205  \\
    2449 & 834 & 0.33125 & 0.01205  \\
    \hline
  \end{tabular}
\end{table}

\begin{table}
  \caption{ The CTPOL parameters. The $a$ and $b$ are parameters in eq. \ref{correction_ct}, $r$ is the cutoff distance. The correction factor $k$ in eq. \ref{correction_ct} is set as 3.418. }
  \label{tbl:CTPOL_params}
  \begin{tabular}{lllll}
    \hline
    Type & Polarizability (nm$^{3}$) & a & b & r (nm)  \\
    \hline
    1142 & 0.002668 & -1.037  & 0.323 & 0.312   \\
    1178 & 0.000729 & -0.246  &  0.072 & 0.294  \\
    1180 & 0.00093 & -0.478 & 0.129 & 0.270 \\
    2178 & 0.000721 & -2.667 & 0.722 & 0.271 \\
    2180 & 0.000901 & -0.635 & 0.172 & 0.270  \\
    2452 & 0.000952 & -0.593 & 0.193 & 0.325  \\
    2444 & 0.000879 & -2.424 & 0.843 & 0.348  \\
    444 & 0.000879 \\
    452 & 0.000952 \\
    834 & 0.004383 \\
    166/2166/1166 & 0.001454 \\
    447/2447 & 0.001341 \\
    448/2448 & 0.001416 \\
    80 & 0.001316 \\
    1177 & 0.001473 \\
    446/2446 & 0.001397 \\
    177/2177 & 0.001441 \\
    178 & 0.000721 \\
    184 & 0.001292 \\
    449/2449 & 0.001446 \\
    180 & 0.000901 \\
    1148 & 0.001475 \\
    96 & 0.000724 \\
    250 & 0.001394 \\
    246 & 0.000906 \\
    235 & 0.001339 \\
    81/82 & 0.001431 \\
    243 & 0.000864 \\
    94 & 0.001457 \\
    108 & 0.001497 \\
    214 & 0.000858 \\
    213 & 0.001685 \\
    230 & 0.000810 \\
    179 & 0.000904 \\
    165 & 0.001425 \\
    90 & 0.001471 \\
    251 & 0.001417 \\
    216 & 0.001504 \\
    109 & 0.000711 \\
    99 & 0.001439 \\
    245 & 0.001410 \\
    \hline
  \end{tabular}
\end{table}

\begin{figure}[htbp]
    \centering
    \includegraphics[width=0.8\textwidth]{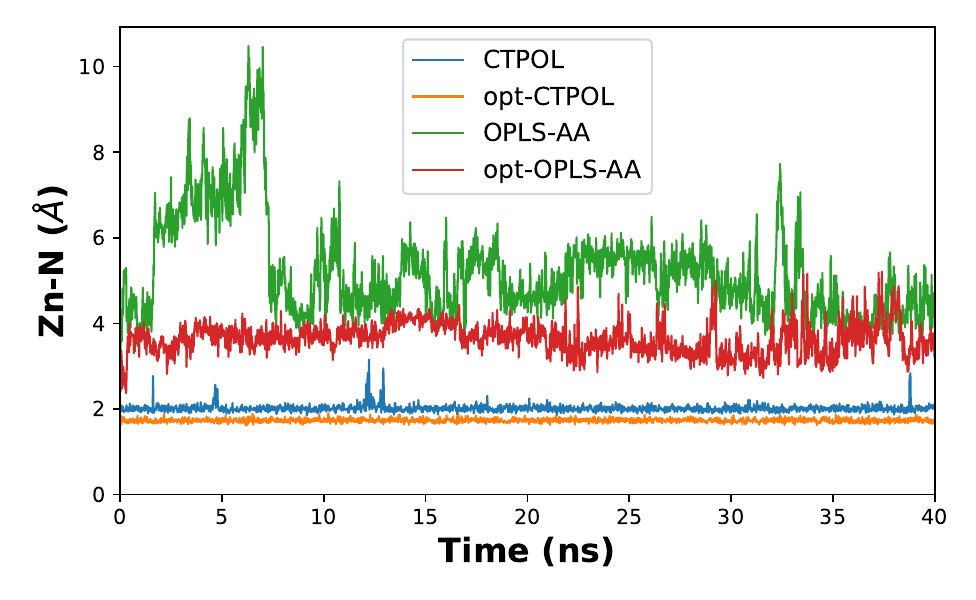}
    \caption{Average of the two Zn-N distances, where the N are the NE2 atoms of the two histidines in the binding site, as a function of time. }
    \label{SI:zn-n_v_time}
\end{figure}

\begin{table}[]
    \centering
 \begin{tabular}{|c|c|c|c|c|}
\hline
PDBid &  Zn\_sites &  N4HC &  N2H2C &  Min H2O dist \\
\hline
 1MEY &         8 &     7 &      7 &          4.38 \\
 4QF3 &         4 &     4 &      0 &          3.98 \\
 6UEI &         4 &     4 &      0 &          4.24 \\
 6UEJ &         4 &     4 &      0 &          4.30 \\
 2PUY &         4 &     4 &      0 &          4.35 \\
 6FI1 &         4 &     4 &      0 &          9.00 \\
 6FHQ &         4 &     4 &      0 &          4.04 \\
 5YC3 &         2 &     2 &      0 &          6.49 \\
 3T7L &         2 &     2 &      0 &          4.41 \\
 3U9G &         4 &     4 &      0 &          4.26 \\
 4Q6F &         8 &     8 &      0 &          4.32 \\
 3IUF &         1 &     1 &      1 &          5.42 \\
 4BBQ &         8 &     8 &      0 &          4.44 \\
 5YC4 &         2 &     2 &      0 &          6.60 \\
 5Y20 &         2 &     2 &      0 &          5.66 \\
\hline
\end{tabular}
    \caption{PDB ids of Zn fingers. 
    N4HC denotes number of Zn binding sites with 4 His and Cys residues, whereas N2H2C denotes the number of Zn binding sites with exactly 2 His and 2 Cys.
    The last column denotes the distance of the closest water molecule to the Zn ion.}
    \label{SI:tbl:other_pdbs}
\end{table}

\begin{figure}

      \includegraphics[width=0.8\textwidth]{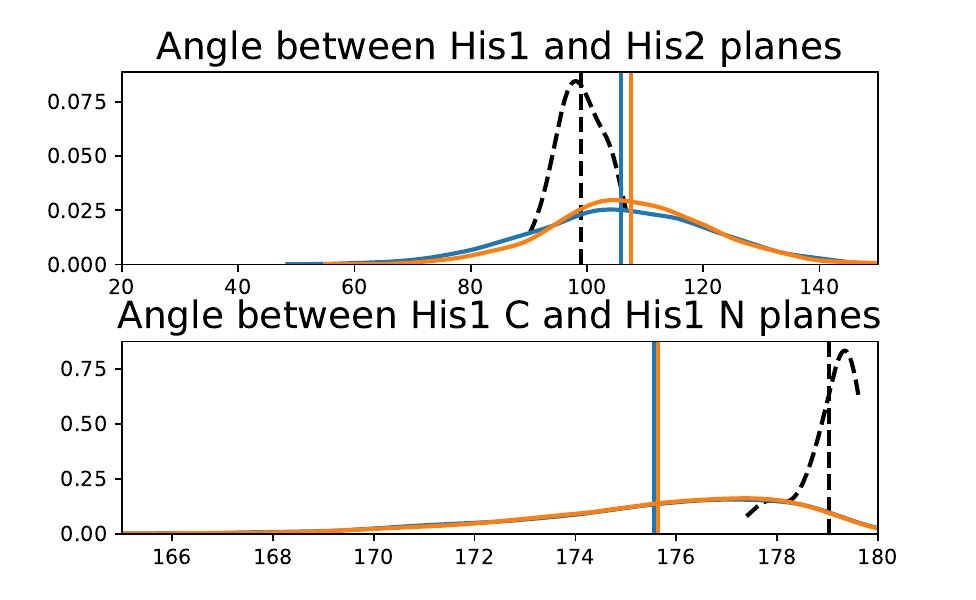}

\caption{ Probability distributions of select dihedral angles. (top) The dihedral angle between the two coordinating histidine planes. 
    The planes were determined using the CG, CD, and CE atoms of histidine. 
    (bottom) The dihedral angle between plane defined by His1  CG, CD, and CE1 atoms, and plane defined by His1 CG, ND, and NE atoms.
    This is to check for internal distortion of the plane.
    The values are close to 180 (instead of 0) because one set of atoms goes clockwise, and the other counter clockwise, when defining the planes.}
    \label{SI:fig:his_ang}
\end{figure}

\end{suppinfo}

\end{document}